\documentclass{aastex63}
\usepackage{lineno}
\usepackage[utf8]{inputenc}
\usepackage[titletoc, title]{appendix}
\usepackage{CJKutf8}
\usepackage{float}
\usepackage{blindtext}
\usepackage{gensymb}
\usepackage{amsmath}
\usepackage{mathtools}

\def \Physics {Department of Physics, University of Michigan, Ann Arbor, MI 48109, USA}
\def \Astronomy {Department of Astronomy, University of Michigan, Ann Arbor, MI 48109, USA}
\def \Washington {Department of Astronomy, University of Washington, Seattle, WA, 98195, USA}
\def \CfA {Center for Astrophysics | Harvard \& Smithsonian, 60 Garden Street, Cambridge, MA 02138, USA}
\def \NAU {Department of Astronomy and Planetary Science, Northern Arizona University,\\PO Box 6010, Flagstaff, AZ 86011, USA}

\graphicspath{{Figures/}}

\begin{document}
\shorttitle{DEEP V: The Absolute Magnitude Distribution of the Cold Classical Kuiper Belt}
\shortauthors{Napier et al.}

\title{The DECam Ecliptic Exploration Project (DEEP): V. The Absolute Magnitude Distribution of the Cold Classical Kuiper Belt}

\correspondingauthor{Kevin J. Napier}
\email{kjnapier@umich.edu}
\author[0000-0003-4827-5049]{Kevin~J.~Napier}
\author[0000-0001-7737-6784]{Hsing~Wen~Lin (\begin{CJK*}{UTF8}{gbsn}
林省文\end{CJK*})}
\affiliation{\Physics}
\author[0000-0001-6942-2736]{David~W.~Gerdes}
\affiliation{\Physics}
\affiliation{\Astronomy}
\author[0000-0002-8167-1767]{Fred~C.~Adams}
\affiliation{\Physics}
\affiliation{\Astronomy}
\author[0000-0001-8994-032X]{Anna~M.~Simpson}
\affiliation{\Physics}
\affiliation{\Astronomy}
\author[0000-0002-9989-4782]{Matthew~W.~Porter}
\affiliation{\Physics}
\author[0000-0002-1435-7102]{Katherine~G.~Weber}
\affiliation{Department of Learning Health Sciences, 
University of Michigan, Ann Arbor, MI 48109, USA}
\author[0000-0002-2486-1118]{Larissa~Markwardt}
\affiliation{\Astronomy}
\author[0000-0002-0644-2956]{Gabriel~Gowman}
\affiliation{\Physics}
\affiliation{\Astronomy}

\author[0000-0002-7895-4344]{Hayden Smotherman} 
\affiliation{\Washington}
\author[0000-0003-0743-9422]{Pedro H. Bernardinelli}
\affiliation{\Washington} 
\author[0000-0003-1996-9252]{Mario Juri\'c} 
\affiliation{\Washington}
\author{Andrew J. Connolly} 
\affiliation{\Washington}
\author{J. Bryce Kalmbach} 
\affiliation{\Washington}
\author{Stephen K. N. Portillo} 
\affiliation{\Washington}

\author[0000-0003-4580-3790]{David E. Trilling}
\affiliation{Department of Astronomy and Planetary Science, Northern Arizona University, Flagstaff, AZ 86011, USA}
\author[0000-0001-6350-807X]{Ryder Strauss}
\affiliation{Department of Astronomy and Planetary Science, Northern Arizona University, Flagstaff, AZ 86011, USA}
\author[0000-0001-5750-4953]{William J. Oldroyd}
\affiliation{Department of Astronomy and Planetary Science, Northern Arizona University, Flagstaff, AZ 86011, USA}
\author[0000-0001-9859-0894]{Chadwick A. Trujillo}
\affiliation{Department of Astronomy and Planetary Science, Northern Arizona University, Flagstaff, AZ 86011, USA}
\author[0000-0001-7335-1715]{Colin Orion Chandler}
\affiliation{Department of Astronomy, University of Washington, Seattle, WA 98195, USA}
\affiliation{LSST Interdisciplinary Network for Collaboration and Computing, Tucson, AZ 85721, USA}
\affiliation{Department of Astronomy and Planetary Science, Northern Arizona University, Flagstaff, AZ 86011, USA}

\author[0000-0002-1139-4880]{Matthew~J.~Holman}
\affiliation{\CfA}

\author[0000-0002-0298-8089]{Hilke E. Schlichting}
\affiliation{Department of Earth, Planetary, and Space Sciences, The University of California, Los Angeles, 595 Charles E. Young Drive East, Los Angeles, CA 90095, USA}

\author{Andrew McNeill}
\affiliation{\NAU}
\affiliation{Department of Physics, Lehigh University, 16 Memorial
Drive East, Bethlehem, PA, 18015, USA}

\author{The DEEP Collaboration}

\begin{abstract}
The DECam Ecliptic Exploration Project (DEEP) is a deep survey of the trans-Neptunian solar system being carried out on the 4-meter Blanco telescope at Cerro Tololo Inter-American Observatory in Chile using the Dark Energy Camera (DECam). By using a shift-and-stack technique to achieve a mean limiting magnitude of $r \sim 26.2$, DEEP achieves an unprecedented combination of survey area and depth, enabling quantitative leaps forward in our understanding of the Kuiper Belt populations. This work reports results from an analysis of twenty 3 sq.\ deg.\ DECam fields along the invariable plane. We characterize the efficiency and false-positive rates for our moving-object detection pipeline, and use this information to construct a Bayesian signal probability for each detected source. This procedure allows us to treat all of our Kuiper Belt Object (KBO) detections statistically, simultaneously accounting for efficiency and false positives. We detect approximately 2300 candidate sources with KBO-like motion at S/N $>6.5$. We use a subset of these objects to compute the luminosity function of the Kuiper Belt as a whole, as well as the Cold Classical (CC) population. We also investigate the absolute magnitude ($H$) distribution of the CCs, and find consistency with both an exponentially tapered power-law, which is predicted by streaming instability models of planetesimal formation, and a rolling power law. Finally, we provide an updated mass estimate for the Cold Classical Kuiper Belt of $M_{\scriptstyle CC}(H_r < 12) = 0.0017^{+0.0010}_{-0.0004} M_{\earth}$, assuming albedo $p = 0.15$ and density $\rho = 1$ g cm$^{-3}$.
\end{abstract}


\accepted{15 September 2023}

\section{Introduction}
\label{sec:intro} 

Beyond the orbits of the major planets, our solar system hosts a large population of minor bodies known as Kuiper Belt Objects (KBOs). In the 30 years since the observational establishment of the Kuiper Belt \citep{jewitt1993}, several surveys (e.g., \citealt{deep-ecliptic-survey, bernstein2004, CFEPSPetit2011, OSSOSVII, bernardinelli2022}) have pushed the inventory of known objects to nearly 4000. These bodies, which are left over from the birth of our planetary system, provide constraints on its formation and dynamical evolution. When taken in aggregate, their dynamics, compositions, and sizes enable us to infer details about the dynamical evolution of the planets, the composition of our solar system's protoplanetary disk, and even the physical processes by which planetesimal formation occurred (see, e.g., \citealt{Nesvorny2018}, \citealt{GladmanVolk2021}).

In particular, the size distribution of the so-called Cold Classicals (CCs)---which are thought to be relics of the birth of the solar system, relatively untouched and uncontaminated in the $\sim4.5$ Gyr since their formation---is a sensitive probe of the process of planetesimal formation. If the CCs are truly a quiescent population, a measurement of their size distribution can provide us with a unique opportunity to directly compare a primordial size distribution with predictions made by planetesimal formation models. Such a comparison will enable us to hone our formation models, and better understand the details of the physical processes at play in planetesimal formation, as well as the specific conditions of our own protoplanetary disk.

Recently, the streaming instability (SI; \citealt{Abod2019}) has begun to emerge as a leading theory of planetesimal formation. Numerical SI simulations predict an exponentially tapered power law absolute magnitude ($H$) distribution, enabling a direct comparison between theory and observation. \citet{Kavelaars2021} found that the absolute magnitude distribution of the CCs detected by the Outer Solar System Origins Survey (OSSOS; \citealt{OSSOSVII}) is consistent with an exponentially tapered power law. However, the CCs used by \citet{Kavelaars2021} only went as faint as $H_r \sim 8.3$, leaving faint-end consistency with the SI as an open question. Existing literature on measurements of the faint end of the CC absolute magnitude distribution seems to be in weak tension with SI models of planetesimal formation. In particular, \citet{fraser2014} find a marginally steeper faint-end size distribution than is predicted by SI simulations. However, a dearth of observed objects at the faint end of the distribution, along with the fact that \citet{fraser2014} did not fit to an exponentially tapered power law, limits the usefulness of such comparisons. We require a larger, deeper set of CC detections by a survey with well-understood biases to thoroughly test any planetesimal formation theory.

The DECam Ecliptic Exploration Project (DEEP) is the first survey with sufficient depth and areal coverage to settle the tension in the shape of the faint end of the $H$ distribution of the CCs. In this paper, we analyze data from 20 DECam fields (comprising an area of approximately 60 sq.~deg.), reaching a mean limiting magnitude of $m_r \sim 26.2$. We use a subset of our data to reconstruct the luminosity function of the Kuiper Belt as a whole, as well as the luminosity function of the CC population. As the main scientific result of this paper, we use our results to reconstruct the underlying absolute magnitude distribution of the CCs, and find consistency with models of planetesimal formation via the streaming instability \citep{Abod2019, Kavelaars2021}.

This paper is organized as follows. In Section~\ref{sec:survey} we outline our observational strategy and discuss the data used in the subsequent analysis. Next we describe our image pre-processing pipeline (Section \ref{sec:images}) and the pipeline used to carry out the object search (Section \ref{sec:pipeline}). We present an overview of our detections in Section \ref{sec:detections}. In Section \ref{sec:efficiency} we calculate our detection efficiency using implanted synthetic objects. We compute the luminosity function for our KBOs as a whole in Section \ref{sec:luminosity}. In Section \ref{sec:dynamics} we isolate a sample of CCs from our detections. Our main scientific result, presented in Section \ref{sec:H}, is a calculation of the absolute magnitude distribution for the CC population. In Section \ref{sec:mass} we calculate an estimate of the total mass of the CCs. In Section \ref{sec:b04} we test the consistency of our absolute magnitude distributions with the results from \citet{bernstein2004}. The paper concludes in Section \ref{sec:conclude} with a summary of our results and a discussion of their implications.

\section{DEEP Survey Strategy and Data}
\label{sec:survey}

DEEP was carried out with the Dark Energy Camera (DECam) on the 4-meter Blanco telescope located at Cerro Tololo Inter-American Observatory in Chile from 2019--2023, targeting four patches of sky along the invariable plane (see \citealt{deepi} and \citealt{deepii}, hereafter Papers I and II, for more details). This paper focuses on the data taken in one of those four patches, our so-called B1 fields, from 2019--2021. These data consist of 20 individual DECam pointings targeting a progressively larger area of sky from 2019--2021, with significant overlap between years in order to enable the tracking of KBOs. Table \ref{tab:B1_fields} gives the pointing for each night of data, as well as the detection efficiency parameters, calculated in Section \ref{sec:efficiency}. Note that in order to avoid double-counting of single-epoch detections we only use a subset of our fields (indicated in Table \ref{tab:B1_fields} to derive the constraints in Sections \ref{sec:luminosity} and \ref{sec:H}.

A DEEP exposure sequence, designed with a cadence ideal for a technique called shift-and-stack \citep{1992AAS...181.0610T,DT1,DT2,bernstein2004,DT4,Parker2010,DT5,kbmod}, typically consists of $\sim$100 consecutive 120-second VR-band exposures of the target field.\footnote{In this work, we treat VR and $r$ equally, as we found that the color term between them was very small. In future work we will measure the VR-$r$ color term and do a proper transformation between the two filters.} In the shift-and-stack approach, single-epoch images are shifted at the rate of a moving object (rather than at the sidereal rate) so that a moving object appears as a point source in the co-added stack.

There are two primary reasons why the shift-and-stack technique is preferable to long exposures for the discovery of moving objects. First, since the rate and direction of an object's motion are not known \textit{a priori}, we are able to stack our images in a grid of velocity and direction that spans the space of possible KBO motions. The second benefit is the preservation of the S/N of moving objects. For stationary sources in astronomical images, S/N goes like $t^{1/2}$. A source is effectively stationary if its apparent position changes by less than the size of the PSF over the course of the exposure. This sets an upper limit on the useful exposure time when searching for moving objects, which we will call $t_{\text{max}}$. Given DECam's typical VR-band seeing of $\sim$0.9'' and the typical KBO rate of motion of a few arcseconds per hour, our value of $t_{\text{max}}$ is on the order of several minutes. When $t \geq t_{\text{max}}$, the S/N of \textit{stationary sources} continues to go like $t^{1/2}$, while smearing causes the S/N of \textit{moving sources} to go like $t^{0}$. Thus moving objects fade into the background while the S/N of background sources continues to grow. Because DECam's CCDs have negligible read noise, we lose no sensitivity by adding together many short images, and thus the S/N of the moving objects continues to increase like $t^{1/2}$.

\begin{figure}[h]
    \centering
    \includegraphics[width=0.32\textwidth]{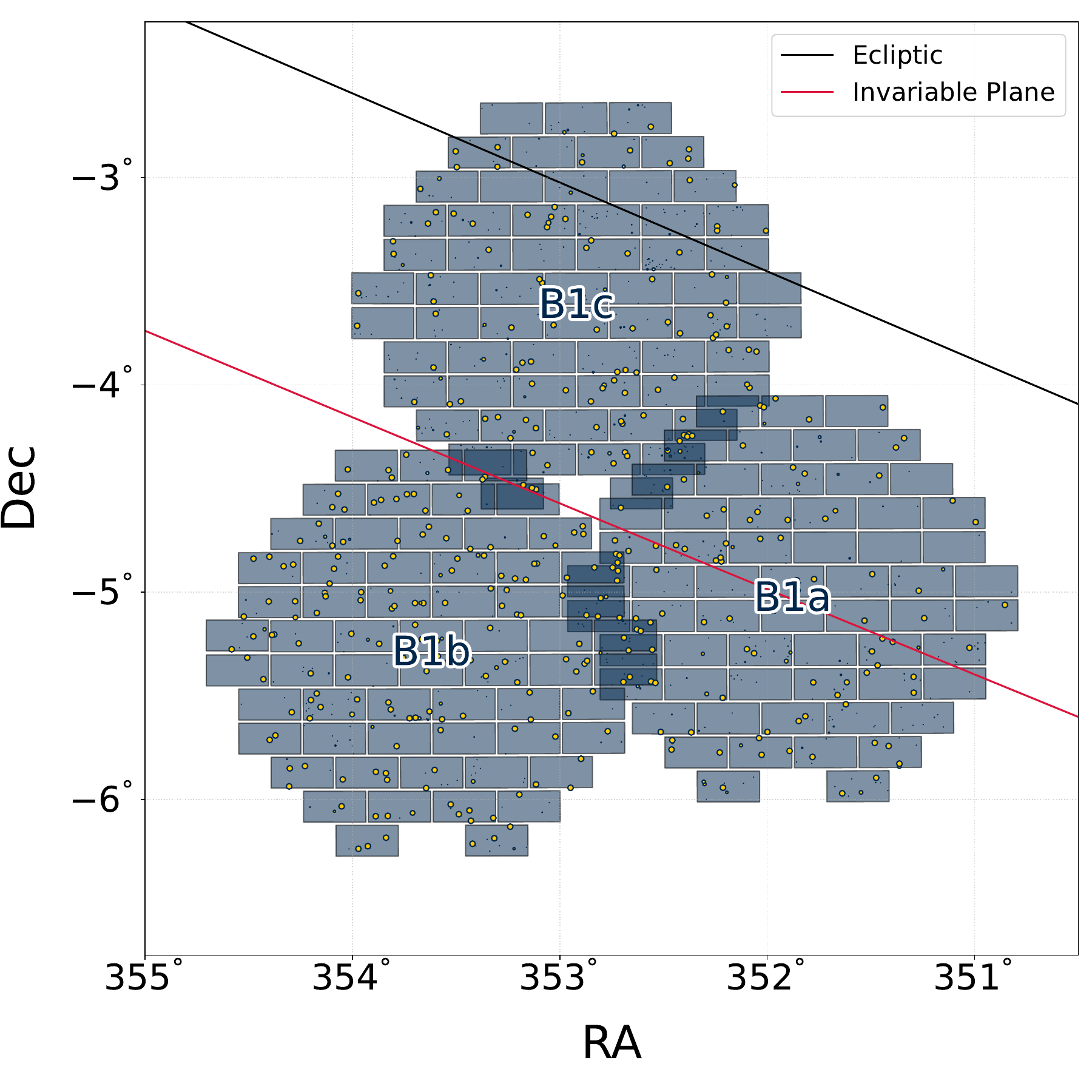}
    \includegraphics[width=0.32\textwidth]{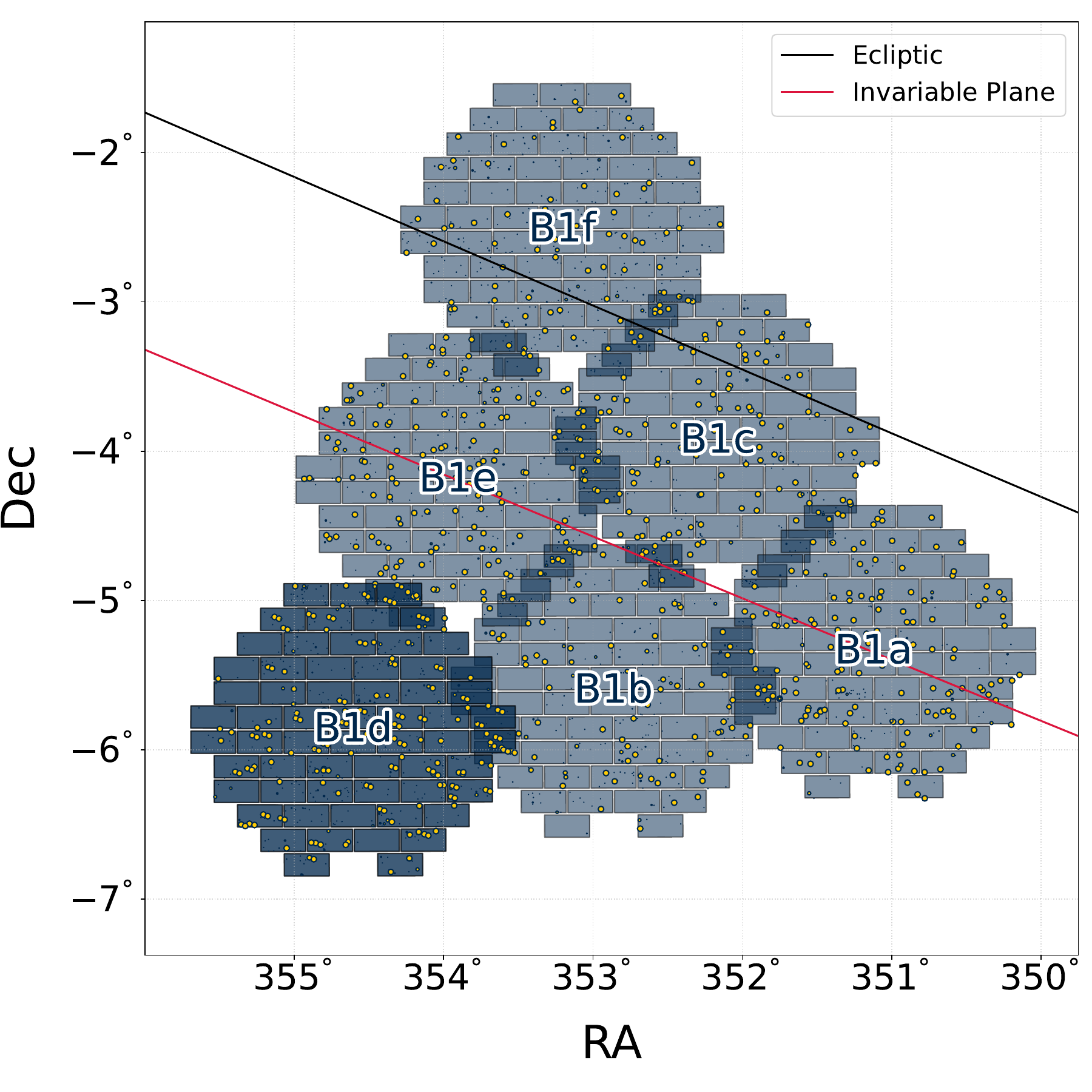}
    \includegraphics[width=0.32\textwidth]{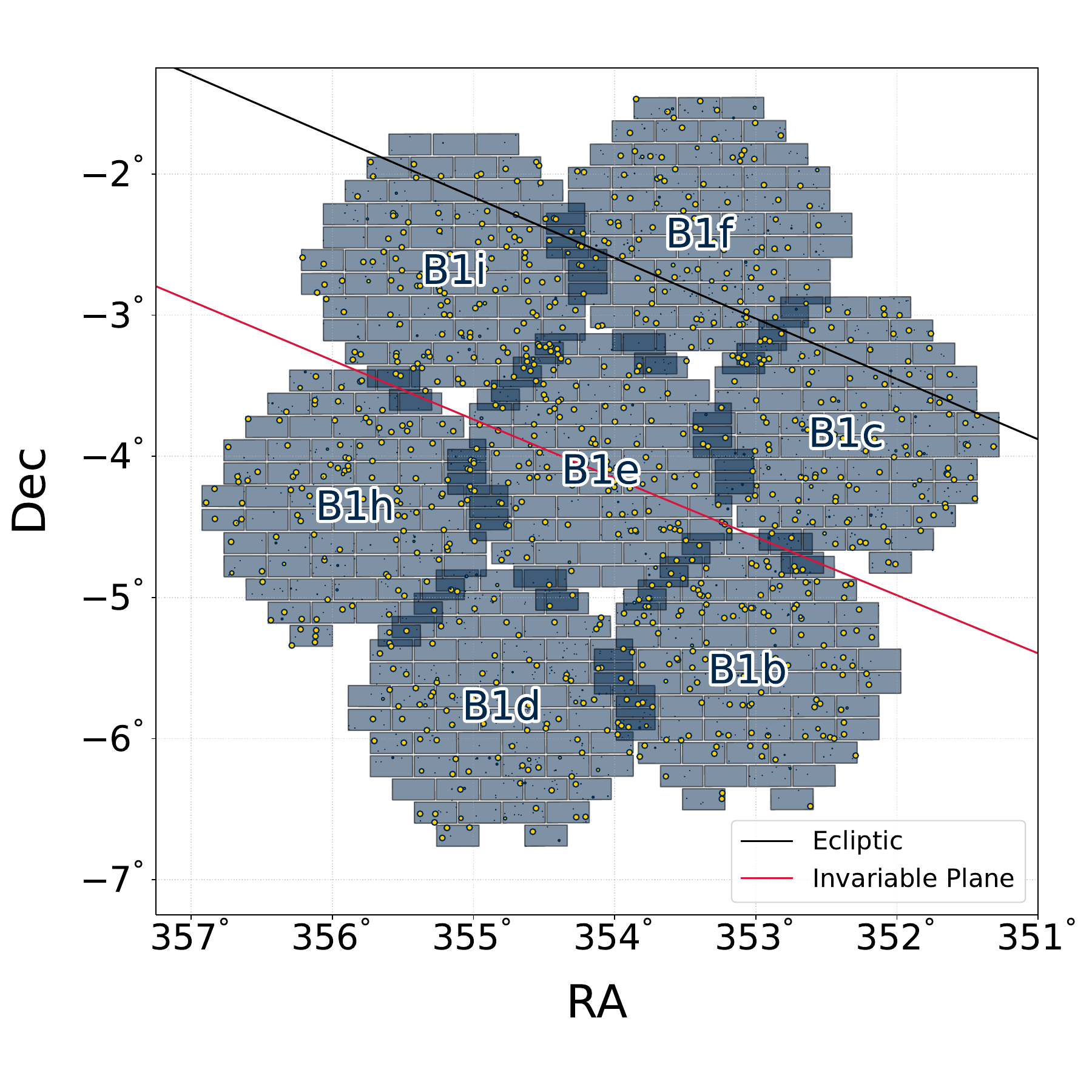}
    \caption{The DEEP ``B1'' TNO search fields used in this analysis. Each hexagonal shaded area represents the DECam focal plane with its 61 active CCDs. The B1a-c fields were observed with integrated exposure times of $\sim 3.5$ hours in August 2019 (left), and re-observed at suitably shifted positions (not plotted) in September 2019. In October 2020 we observed the B1a-f fields (center). In October 2021 we observed the B1b-f, h, and i fields. The larger areas in 2020 and 2021 account for diffusion of the 2019 detections. The plotted points represent our putative KBO detections. Note that Neptune is near the B1 fields, meaning that there should be relatively few resonant objects among our detections.}
    \label{fig:B1_fields}
\end{figure}

\begin{table}[H]
\renewcommand{\arraystretch}{0.95}
\centering
\begin{tabular*}{1.0\textwidth}{l@{\extracolsep{\fill}}lcccc|ccc|cc}
\hline
\hline
Night & Field & JD & RA (J2000) & DEC (J2000) & $N_{exps}$ & $\eta_0$ & $m_{50}$ & $\sigma$ & KBO fits & CC fits \\ \hline
    2019-08-27 & B1c & 2458723.710 & 352.92015 & -3.62314 & 103 & 0.94 & 26.11 & 0.31 & no & no \\ 
    2019-08-28 & B1a & 2458724.711 & 351.87745 & -5.03572 & 102 & 0.95 & 26.08 & 0.32 & no & no \\
    2019-08-29 & B1b & 2458725.711 & 353.61898 & -5.29744 & 101 & 0.93 & 26.65 & 0.31 & no & no \\ 
    \hline
    2019-09-26 & B1a & 2458753.513 & 351.38115 & -5.24000 & 95  & 0.91 & 26.02 & 0.36 & no & no \\
    2019-09-27 & B1b & 2458754.511 & 353.12082 & -5.50267 & 97  & 0.94 & 26.09 & 0.33 & no & no \\
    2019-09-28 & B1c & 2458755.513 & 352.42444 & -3.82797 & 96  & 0.92 & 26.42 & 0.36 & no & no \\ 
    \hline 
    2020-10-15 & B1b & 2459138.499 & 352.86350 & -5.60964 & 103 & 0.88 & 26.22 & 0.29 & yes & yes \\ 
    2020-10-16 & B1c & 2459139.510 & 352.16593 & -3.93469 & 96  & 0.90 & 26.47 & 0.35 & yes & yes \\
    2020-10-17 & B1e & 2459140.511 & 353.90520 & -4.19542 & 96  & 0.90 & 26.23 & 0.34 & yes & yes \\
    2020-10-18 & B1a & 2459141.514 & 351.12145 & -5.34617 & 91  & 0.90 & 26.57 & 0.39 & yes & yes \\
    2020-10-19 & B1d & 2459142.503 & 354.60670 & -5.86869 & 99  & 0.92 & 25.99 & 0.32 & no & no \\
    2020-10-20 & B1f & 2459143.503 & 353.20607 & -2.52153 & 91  & 0.93 & 25.92 & 0.30 & yes & no \\
    2020-10-21 & B1d & 2459144.503 & 354.60635 & -5.86972 & 99  & 0.92 & 26.05 & 0.36 & yes & yes \\ 
    \hline
    2021-09-27 & B1d & 2459485.512 & 354.80112 & -5.78794 & 89  & 0.93 & 25.96 & 0.36 & no & no \\
    2021-10-01 & B1b & 2459489.526 & 353.05852 & -5.52967 & 81  & 0.90 & 26.10 & 0.29 & no & no \\
    2021-10-02 & B1f & 2459490.511 & 353.40057 & -2.44119 & 89  & 0.94 & 26.01 & 0.52 & no & no \\
    2021-10-03 & B1i & 2459491.501 & 355.13703 & -2.69892 & 100 & 0.93 & 26.26 & 0.30 & no & no \\
    2021-10-04 & B1c & 2459492.499 & 352.36053 & -3.85436 & 102 & 0.92 & 26.12 & 0.33 & no & no \\
    2021-10-05 & B1h & 2459493.502 & 355.83970 & -4.37136 & 97  & 0.92 & 26.21 & 0.33 & no & no \\
    2021-10-06 & B1e & 2459494.504 & 354.09928 & -4.11514 & 97  & 0.85 & 26.08 & 0.27 & no & no \\
    \hline
\end{tabular*}
\caption{DEEP telescope pointings used for the long-stare image sequences described in this analysis. The positions of the fields at each epoch aim to track as many objects as possible by accounting for the effects of Earth reflex motion and TNO shear. Each exposure is 120 seconds long and is taken in the VR band. The next three columns show the best fit of each night to Equation (\ref{eq:efficiency}). The final two columns indicate whether a field was used to derive constraints for either the full KBO population or the CC population.}
\label{tab:B1_fields}
\end{table}

\section{Image Pre-processing}
\label{sec:images} 

In this section we describe how our images are processed in preparation for our shift-and-stack pipeline (described in Section \ref{sec:pipeline}). The following steps take place after the images have gone through preliminary reductions with the DECam community pipeline \citep{CommunityPipeline}.

\subsection{Synthetic TNOs}
\label{sec:fakes} 
To enable studies of our efficiencies, we generated a population of several thousand synthetic sources to plant in our images. These synthetic sources were not meant to emulate a realistic population, but rather to test efficiency across the space of all possible bound orbits in the Kuiper Belt. They span distances from $\sim20$ au to a few hundred au, and include fully retrograde orbits. To enable studies of efficiency as a function of brightness, we have given our synthetic sources apparent magnitudes as bright as 20, and as faint as 27.2, as well as sinusoidal rotation curves with amplitudes as large as 0.5 mag, and rotation periods between a few hours and a few days. 

\subsection{Flux Calibration and Synthetic Source Injection}

To calibrate the flux of our synthetic sources we calculate the photometric zero-point for each individual CCD image by cross-matching the non-streaked sources (ellipticity $< 0.8$\footnote{Most of the images can successfully match with enough sources with smaller ellipticity. However, a small number of them may fail due to some issues with the images, e.g. bad telescope guiding. This value ensures that every image can be processed using the same pipeline parameter settings. Since galaxies may also be included during this process, we clip photometric outliers to purify the sources for calibration.}) against Pan-STARRS sources \citep{PS1_Photometry2} with r$_{SDSS}$ magnitude\footnote{The r$_{SDSS}$ magnitude were converted from the Pan-STARRS r$_{P1}$ and g$_{P1}$ magnitude using \citet{PS1_Photometry1}.} between 15 and 21. We then use the Python package {\tt SpaceRocks} \citep{Napier_spacerock} to calculate the sky position, sky motion, and brightness of each synthetic TNO, including a rotation lightcurve. With the sky motion of each object and the PSF of the image, we generate a streak model for each synthetic TNO. With the brightness, streak model, and photometric zero-point specified, we inject the synthetic TNOs into the image.\footnote{This step is performed prior to template construction.} Along with the synthetic TNOs, we also inject 12 stationary synthetic point sources with an r-band magnitude of 21 into each CCD image, in order to enable calibration after difference imaging. We add these 12 stationary sources at fixed pixel locations in the images (i.e., not fixed sky positions), so they do not appear in the template, and thus remain in difference images.

\subsection{Difference Imaging}

After we implant synthetic sources, we prepare the images for the shift-and-stack pipeline. To do this, we must remove every stationary source---even the faintest sources that are not visible in single exposures. We apply the High Order Transform of Psf ANd Template Subtraction code \textsc{hotpants} \citep{hotpants}, which implements and improves upon the method of \citet{Alard1998} to create difference images. This code formed the basis for the Dark Energy Survey's supernova search pipeline \citep{Kessler2015}, and has consequently been thoroughly exercised on DECam data.

We first assemble the collection of exposures in the same observing run, including images from both the long and short stares (see Papers I and II for details). We generate three different templates by median-combining the single epoch images with seeing (by measuring the FWHM of the in-frame stars) in the top, middle, and bottom 1/3 of the ensemble. We require the minimum time separation between observations to be longer than 0.01 days (14.4 minutes) to ensure that the templates contain minimal flux from the slow movers.
The \textsc{hotpants} algorithm then performs seeing matches between science images and the template with the closest match to the image's seeing to generate difference images. The better-seeing images (either single epoch or template) are convolved to match the images with the worst seeing, and the bright sources (pixel counts $> 3000$) are masked before performing image subtraction. The final step is masking the bright Gaia sources (G $>18$) and regions where there is a contiguous group of at least 5 pixels above/under the $\pm 2\sigma$ level. This step usually masks less than 1\% of the total pixels and not only cleans out most of the artifacts generated by the difference imaging process, but also removes streaks from artificial satellites, thus greatly reducing the false detection rate in shift-and-stack images. However, this masking comes with the caveat of masking bright sources. In practice, we find that sources brighter than VR$\sim 24$--$24.5$ are masked. To ensure that we recover the bright objects, we also write out difference images in which we do not mask the pixels above/under the $\pm 2\sigma$ level. While the un-masked images produce significantly more spurious sources after shifting and stacking, we simply ignore the faint sources produced by these images, opting to only consider sources with S/N$\geq 30$, thus finding all of the bright sources with minimal additional effort.

Finally we use the stationary magnitude 21 fakes to re-scale each difference image such that it has a zero-point of 30. After re-scaling, we compute weight images as the inverse of the variance in the difference images. Using weight images enables us to optimize the S/N of our stacks without manually rejecting images.\footnote{The weighting isn't quite optimal because variance of the difference images is suppressed by the smoothing that occurs from interpolating the images onto a grid, and because the seeing differs from exposure to exposure, but it works well.}

\section{The Detection Pipeline}
\label{sec:pipeline} 

\subsection{The Grid}
\label{sec:grid}


To carry out our search, we developed a novel method of generating the shift-and-stack grid. We begin by computing the grid bounds. A unique grid is required for each field, as its exact shape depends on the epoch and sky position of a pointing. We must also select the range of topocentric distances ($\Delta$) of interest. For the search described in this paper, we consider $\Delta \in [35, 1000]$ au.\footnote{Note that since we are searching through retrograde orbits, we are sensitive to objects at closer distances.} With RA, Dec, and $\Delta$ fixed, an object's position vector in the topocentric frame, $\vec{x}_T$ is uniquely determined. We then change the origin from the topocentric to the barycentric frame so that we have $\vec{x}_B$.\footnote{For bodies in the inner solar system one should transform to the heliocentric frame, though in practice the distinction makes little difference.} After changing the origin to the barycentric frame, we assign a velocity $v_{\text{bound}}$ such that an object at the position $\vec{x}_B$ would be barely bound to the solar system (specifically, we use the speed appropriate for a semi-major axis $a$ = 200,000 au). We then uniformly sample a collection of velocity vectors $\vec{v}_B$ on the surface of the sphere of radius $v_{\text{bound}}$. With the velocity vectors specified, the state vectors are fully determined, allowing for the computation of the corresponding RA rates and Dec rates as observed in the topocentric frame.\footnote{At this point, one can make cuts on the objects to disallow certain kinds of orbits as a function of Keplerian elements. For example, making a cut to consider only prograde orbits will drastically reduce the size of the grid.} We repeat this process at several discrete values of $\Delta$, and then use the Python package \texttt{alphashape} to draw a concave hull bounding the computed rates. This hull encompasses the full region of physically possible sky motions for the distances of interest.

\begin{figure}[h]
    \centering
    \includegraphics[width=0.8\textwidth]{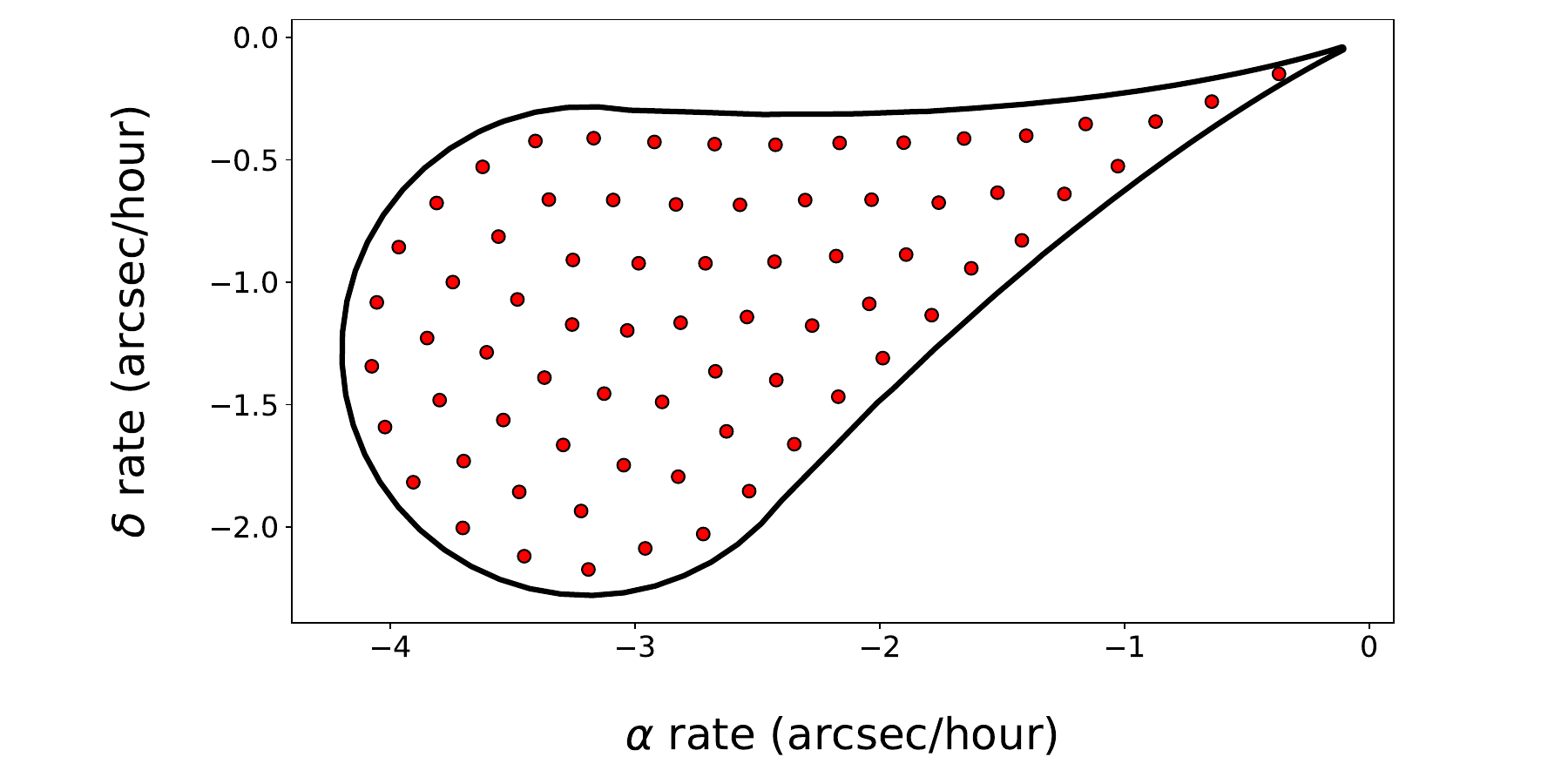}
    \caption{Sample grid for a 4-hour exposure sequence with 1'' seeing. The black teardrop-shaped boundary encompasses the space of possible bound orbits with topocentric distance between 35 and 1000 au. The red points are the sample rates computed using the procedure described above.}
    \label{fig:rates-grid}
\end{figure}

Once we have computed the boundary of the concave hull for a given pointing, we choose a finite set of rates at which to stack our data. Toward this end, we employ a new method in which we fill the hull with a large sample of random points, and then use a K-means clustering algorithm to divide the region into $N$ clusters. One can use dimensional analysis to determine that $N \propto A_{\text{hull}} / \epsilon^2$, where $A_{\text{hull}}$ is the area of the hull, and $\epsilon$ is the desired grid spacing to minimize the maximum source trailing (roughly determined by the PSF width divided by the duration of the exposure sequence). We take the centroid of each cluster as a grid point. When compared to a rectangular grid, this method allows us to use $\sim$20\% fewer grid points, and simultaneously reduce both the mean and maximum distance of any given point in the hull to the nearest grid point. As a result, we achieve more even coverage of physically possible rates, while minimizing the computational cost and opportunities for false positives. Although the process is random by nature, a sufficient number of random samples makes it nearly deterministic. We show an example of a grid for a 4-hour exposure sequence with 1'' seeing in Figure \ref{fig:rates-grid}.

\subsection{The Shift-and-Stack Procedure}
\label{sec:shiftstack} 

After computing the shift-and-stack rates, we proceed with the stacking. We designate the first exposure in an exposure sequence as a reference image. We then compute the RA and Dec at the center of the reference image, and use the RA rate and Dec rate to compute the amount by which we have to shift each image to match with the reference image's center. In other words, we stack the pixels along the path taken by the center of the reference image for given RA and Dec rates. We do not consider variations in focal plane geometry across the chip, as the solid angle of a single chip is rather small (we thus assume that the chips are locally flat), and all of the images are \texttt{SWarped} \citep{SWarp} for the difference imaging process. We perform separate stacks for both the weighted signal (i.e. the signal multiplied by the weight) and weight images, and then obtain the full stack by dividing the stacked weighted signal image by the stacked weight image.

After each stack we use the Python package \texttt{sep} \citep{sep, SourceExtractor} to extract all sources with at least 3 pixels with values above 1.5$\sigma$.\footnote{For completeness, we note that we used the \texttt{matched} filter for our detections. However, the details are of little consequence, as even a simple peak finding algorithm works well for this purpose.} Each stack is contaminated with of order a few $\times 10^3$ spurious sources mostly consisting of cosmic rays, dead pixels, over-saturated pixels close to bright stars, and residuals from poorly subtracted stars and galaxies. Since we use approximately 100 stacks per chip, the total number of spurious sources per chip is close to $10^5$. Because the vast majority of spurious sources are not PSF-like, we have trained convolutional neural nets (CNNs) using \texttt{tensorflow} \citep{tensorflow} to reject them automatically. We trained one CNN on synthetic sources superimposed on background from DEEP difference images, and another on the \texttt{autoscan} training set that was used to train a random forest algorithm for background rejection in the Dark Energy Survey (DES) supernova search \citep{autoscan}. Both CNNs retain nearly all of the signal and fail to reject different types of background, thus enabling a significant performance gain by requiring a source to be classified as \textit{good} by both CNNs. This procedure cuts the number of sources per stack by three orders of magnitude, down to a few $\times 10^2$. After all of the stacks are completed for a given chip, we consider the complication that most objects are bright enough to be detected in adjacent stack rates. To eliminate this redundancy, we employ a \texttt{DBSCAN} \citep{dbscan} clustering algorithm in pixel space to group detections associated with the same object.

\begin{figure}[h]
    \centering
    \includegraphics[width=0.9\textwidth]{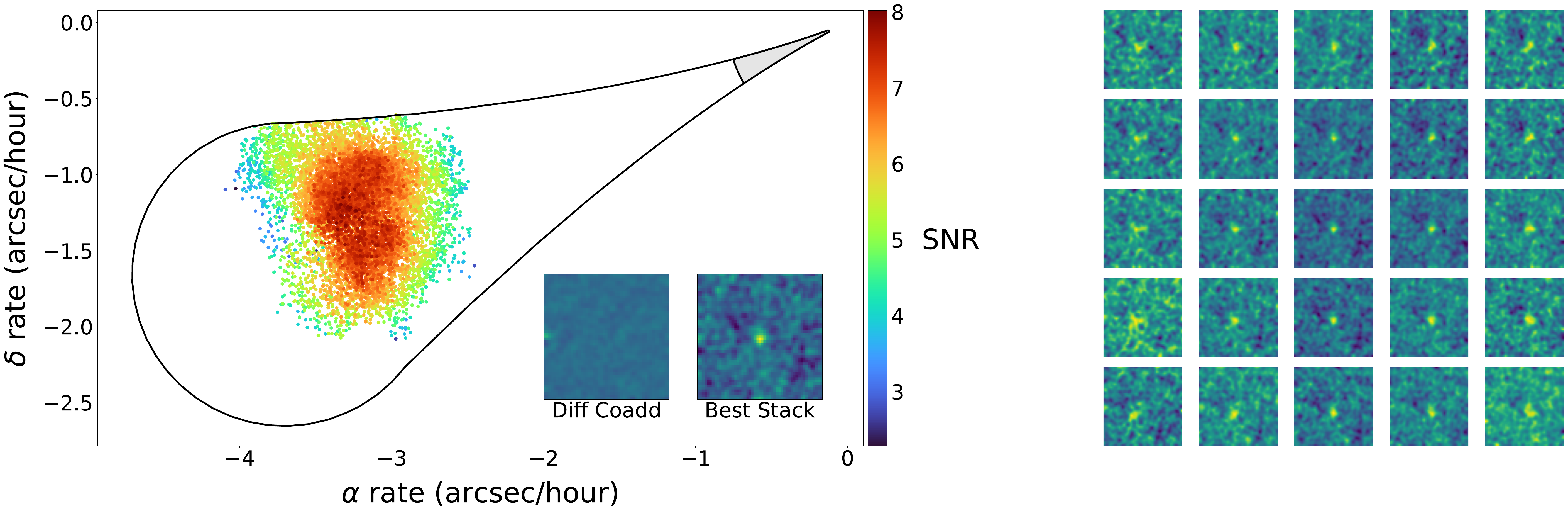}
    \caption{Example of the image format used for visual inspection of our candidates. The detection shown in this figure is a synthetic source with an r-band magnitude of 27.0. The left panel is what we call an MCMC plot. The black teardrop-shaped region is the boundary of the space of possible KBO rates of motion. The grey region at slow rates represents the rate cut we made on our detections. The points are sampled from our MCMC, with colors corresponding to the S/N of the object in the stack (this object had a peak S/N of 8). Each point represents a targeted stack at the given rate. As a stack approaches the correct rate, causing the source to appear optimally point-like, the S/N increases in this characteristic manner. The inset labeled \textit{Diff Coadd} is the stationary stack; the image shows no discernible signal. The inset labeled \textit{Best Stack} is the stack at the best rate, as determined by the MCMC. In this image the KBO is quite apparent. The right panel shows a grid of stacks centered at the best rate, and offset in increments of 1 pixel per hour in RA rate and Dec rate.}
    \label{fig:grid}
\end{figure}

The grid spacing in the initial shift-and-stack is good enough for source detection, but is too coarse to provide the best values for the position and rate for a given source. To refine the parameters, we use a Markov Chain Monte Carlo (MCMC) approach in which we perform targeted stacks on our candidates to maximize S/N. These targeted stacks are still restricted to the parameter space of bound orbits, but are now continuous in RA and Dec rates. This procedure enables us to obtain refined RA and Dec rates with uncertainties, while simultaneously optimizing the measured RA and Dec of the source. We use the uncertainties in RA rate and Dec rate (which are typically about 0.1''/hour in each dimension) to probabilistically classify our detections in Section \ref{sec:dynamics}. After refining the parameters of our detections, we discard all sources with rates slower than 3 pixels per hour (0.79''/hour, or distance $\gtrsim$ 150 au), as such slow rates tend to accumulate false positives due to subtraction artifacts much more quickly than faster rates. We feed our remaining candidates through a final CNN that reduces the number of sources per chip to $\sim 10$. The images we show to the CNN are similar to the right panel in Figure \ref{fig:grid}, and contain more information than the cutouts we show to the first CNN. Good sources tend to show a characteristic radial pattern, while false sources do not. 

Once we have refined our sources' rates and positions, we compute their flux and flux uncertainty using \texttt{sep}. We use these values to calibrate the magnitudes of our detections against the known magnitudes of our implanted sources, as well as obtaining a magnitude uncertainty.

\begin{figure}[H]
    \centering
    \includegraphics[width=0.9\textwidth]{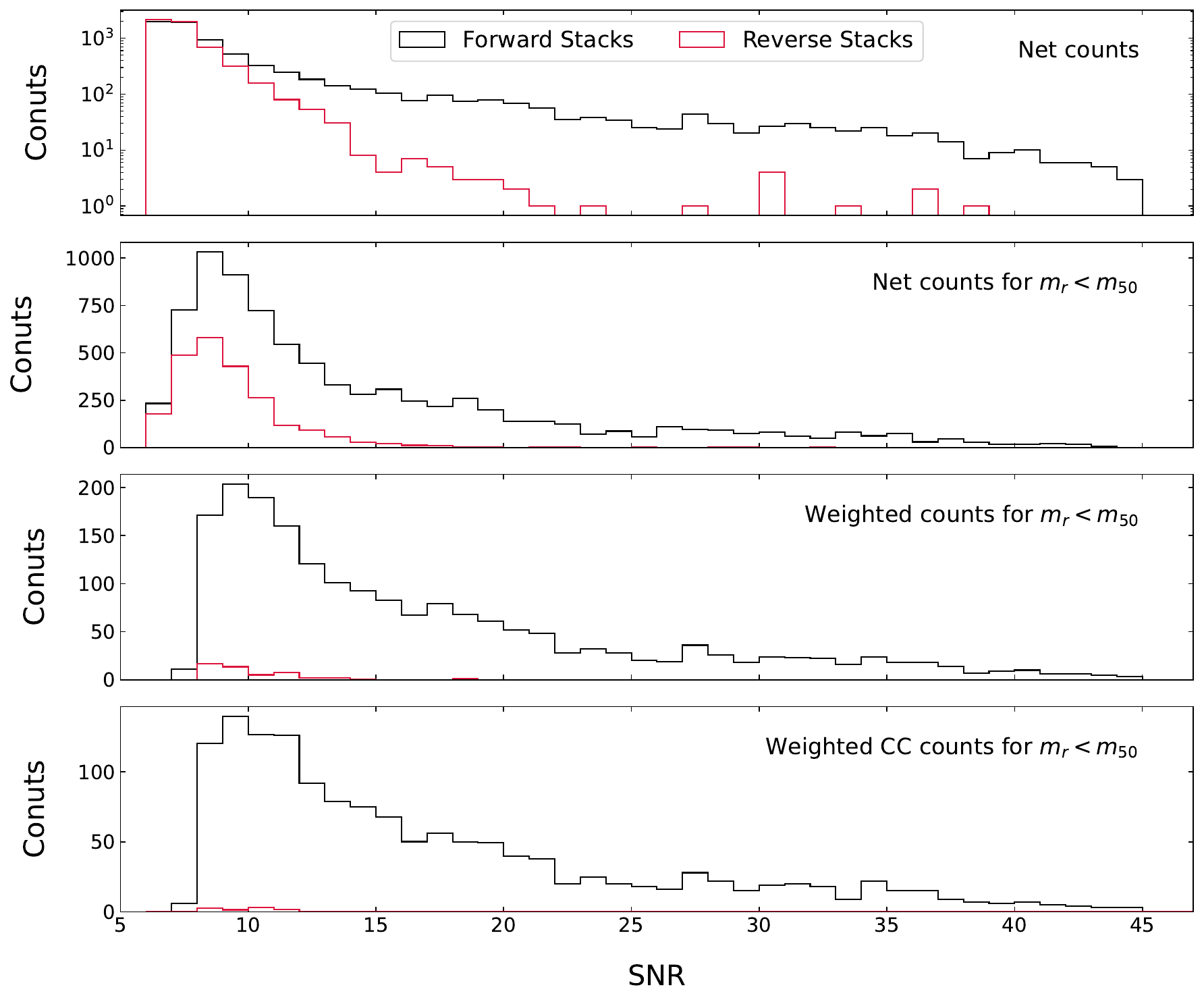}
    \caption{Number of sources resulting from the forward and reverse shift-and-stack as a function of S/N, both prior to human vetting, and after applying weights from human vetting and various cuts. It is clear that false positives drop off quickly with increasing S/N, and that our vetting procedure was highly effective at removing false positives.}
    \label{fig:counts}
\end{figure}

\begin{figure}[H]
    \centering
    \includegraphics[width=0.9\textwidth]{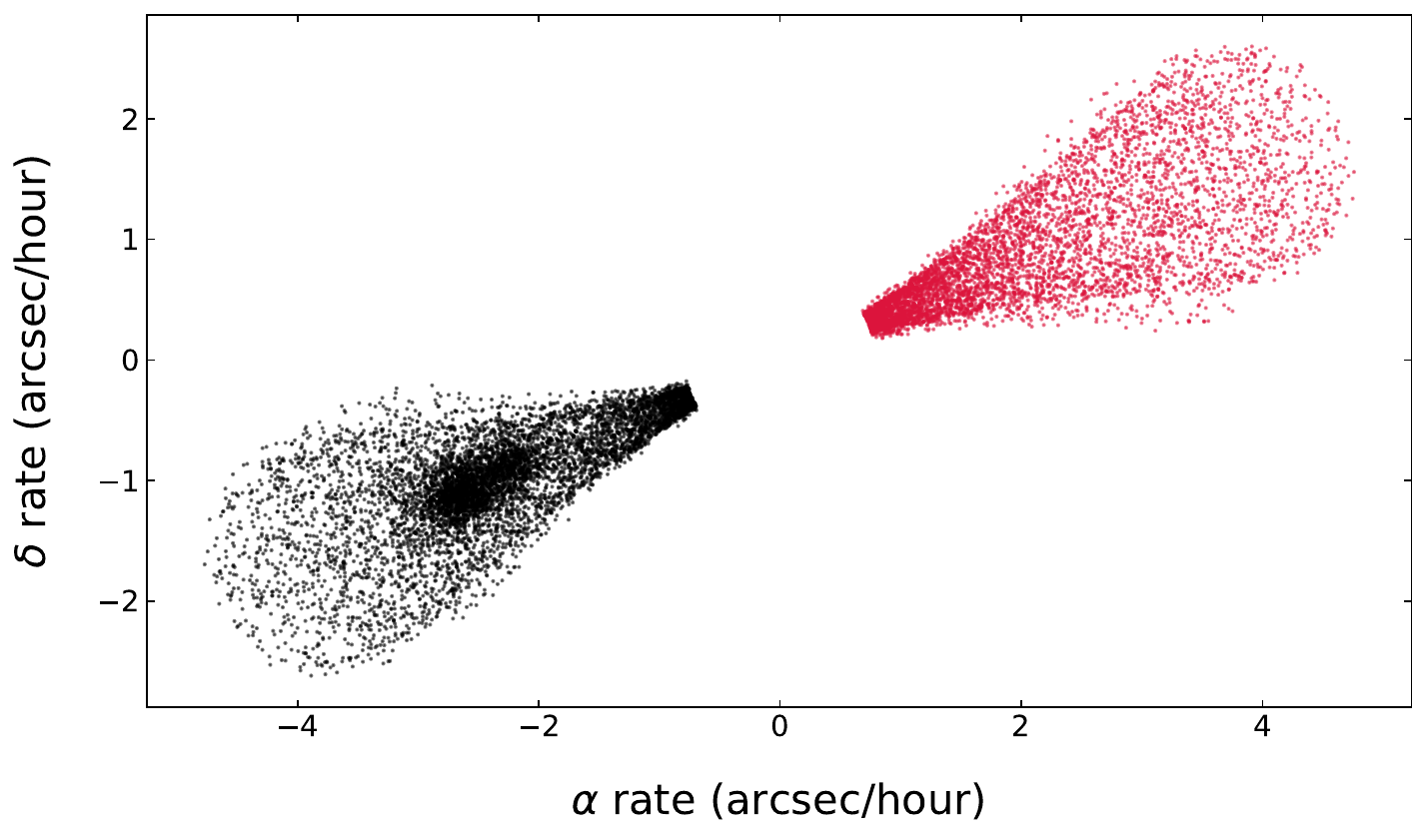}
    \caption{Rate vectors for all detections resulting from the forward (black) and reverse (red) shift-and-stack, prior to human vetting. There is a clear excess of detections in the forward shift-and-stack, corresponding to the Kuiper Belt.}
    \label{fig:rates}
\end{figure}

Finally we do a reverse stack on our data, in which we repeat the above procedure with negated RA and Dec rates. Because no physical KBO would appear as a point source when stacked at these rates, all sources that result from this stack are false positives. This reverse stack enables the critical step of accounting for false positives in our detections. In Figure \ref{fig:counts} we show differential histograms of the number of sources resulting from the forward and reverse shift-and-stack as a function of S/N, both before and after applying weights and various cuts. In Figure \ref{fig:rates} we show a scatter plot of the RA and Dec rates of all detections from the forward and reverse shift-and-stack, prior to human vetting.

\subsection{Candidate Vetting}
The final step in the discovery pipeline is the visual inspection of the grids that were labeled as \textit{good} by the final CNN. This step includes the implanted synthetic sources and the sources from the reverse shift-and-stack. To do this visual inspection we developed a website where users vote \texttt{yes}, \texttt{no}, or \texttt{maybe} on a candidate and have their vote recorded to a database. The images are presented to the vetters in a blind manner, meaning that the vetter has no indication of whether an object is an implanted source, a source from the reverse shift-and-stack, or a true candidate. By blindly vetting all sources, we can reliably compute a voter's true and false positive rates for \texttt{yes}, \texttt{no}, and \texttt{maybe} votes. We require votes from three unique vetters for each object, and then combine the votes into a probability that a source is ``real" using the following framework.

The odds (i.e., betting odds) of a source being good given a vote are
\begin{equation}
    O(+|\text{vote}) = O(+)\frac{P(\text{vote}|+)}{P(\text{vote}|-)}
    \label{eq:odds}
\end{equation}
where $O$ represents odds and $P$ represents probability. The $+$ symbol means that the object is truly a good source, and the $-$ symbol means that the object is truly a bad source. We calculate the prior $O(+)$ using the excess in the number of sources in the forward shift-and-stack over the number of sources in the reverse shift-and-stack. The fraction on the right-hand-side of Equation (\ref{eq:odds}) is the Bayes factor,
\begin{equation}
    B \equiv \frac{P(\text{vote}|+)}{P(\text{vote}|-)}.
\end{equation}
The quantity $P(\text{vote}|+)$ is calculated as the probability of assigning a given vote to an implanted source. Similarly, the quantity $P(\text{vote}|-)$ is calculated as the probability assigning a given vote to a source from the reverse shift-and-stack. 

Given multiple votes, we can simply update the information by taking the product of Bayes factors as
\begin{equation}
    O(+|\text{votes}) = O(+) \prod_i B_i
    \label{eq:bayesproduct}
\end{equation}
where $B_i$ is the Bayes factor of the $i$th voter. We can then convert the odds from Equation (\ref{eq:bayesproduct}) into the probability that a source is real as 
\begin{equation}
    P(+|\text{votes}) = \frac{O(+|\text{votes})}{1 + O(+|\text{votes})}.
    \label{eq:bayes-probability}
\end{equation}
We assign the values calculated by Equation (\ref{eq:bayes-probability}) as a weight ($w$) for each of our detections. This treatment allows us to take a probabilistic approach in studying our detections in Sections \ref{sec:efficiency}-\ref{sec:H}.\footnote{It is only valid to treat these weights as probabilities if the prior expectation agrees reasonably well with the posterior, which was true in this work.}

\begin{figure}[H]
    \centering
    \includegraphics[width=0.49\textwidth]{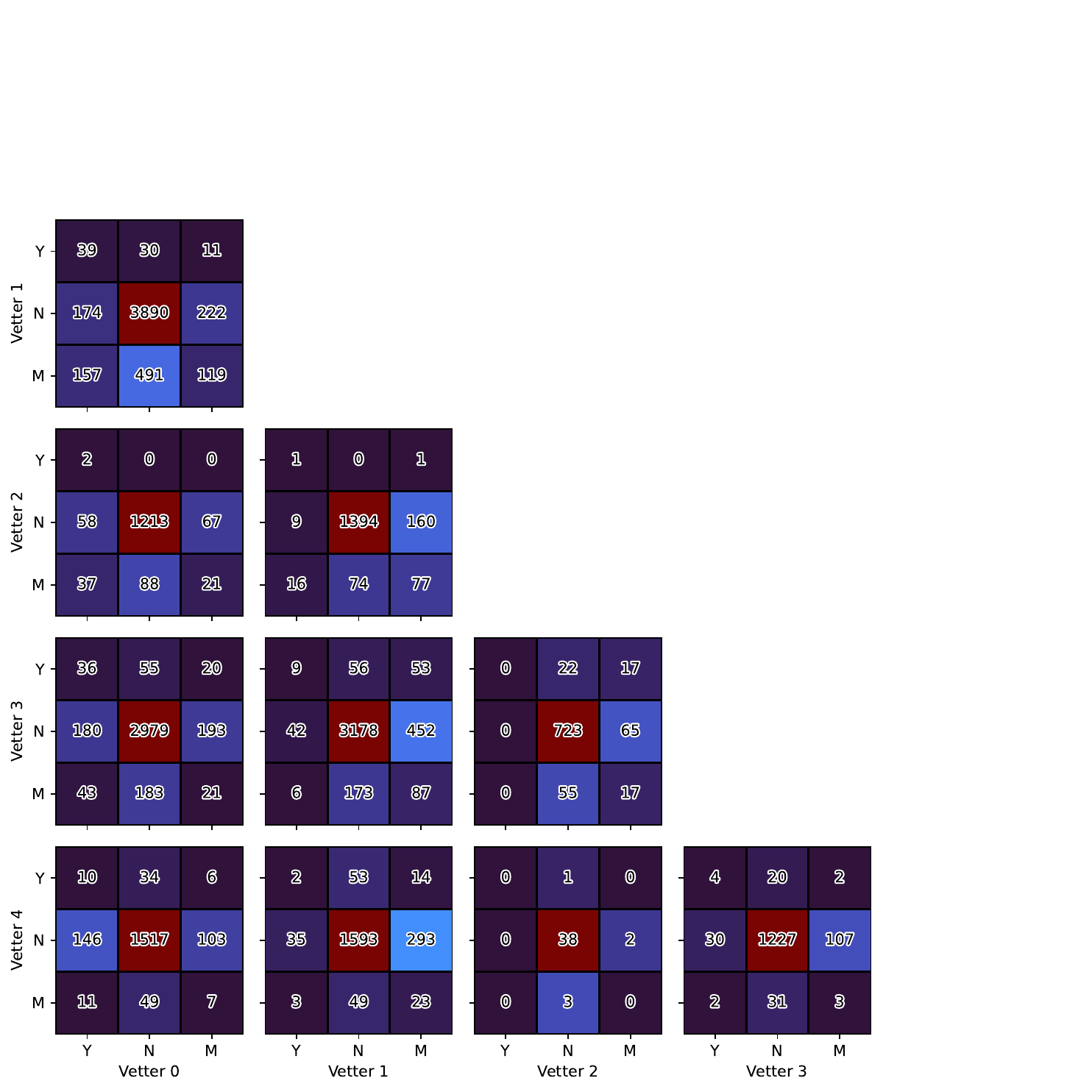}
    \includegraphics[width=0.49\textwidth]{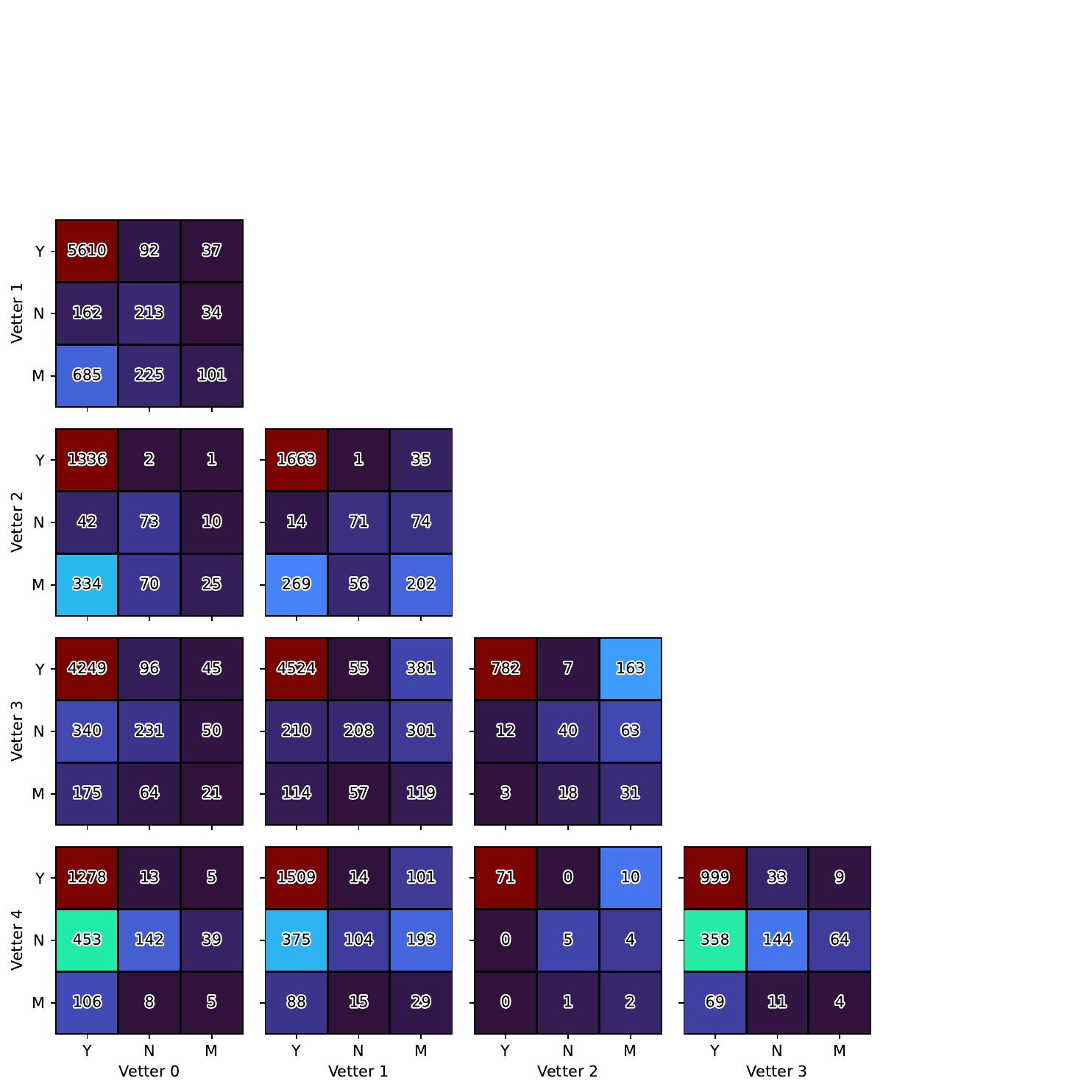}
    \includegraphics[width=0.49\textwidth]{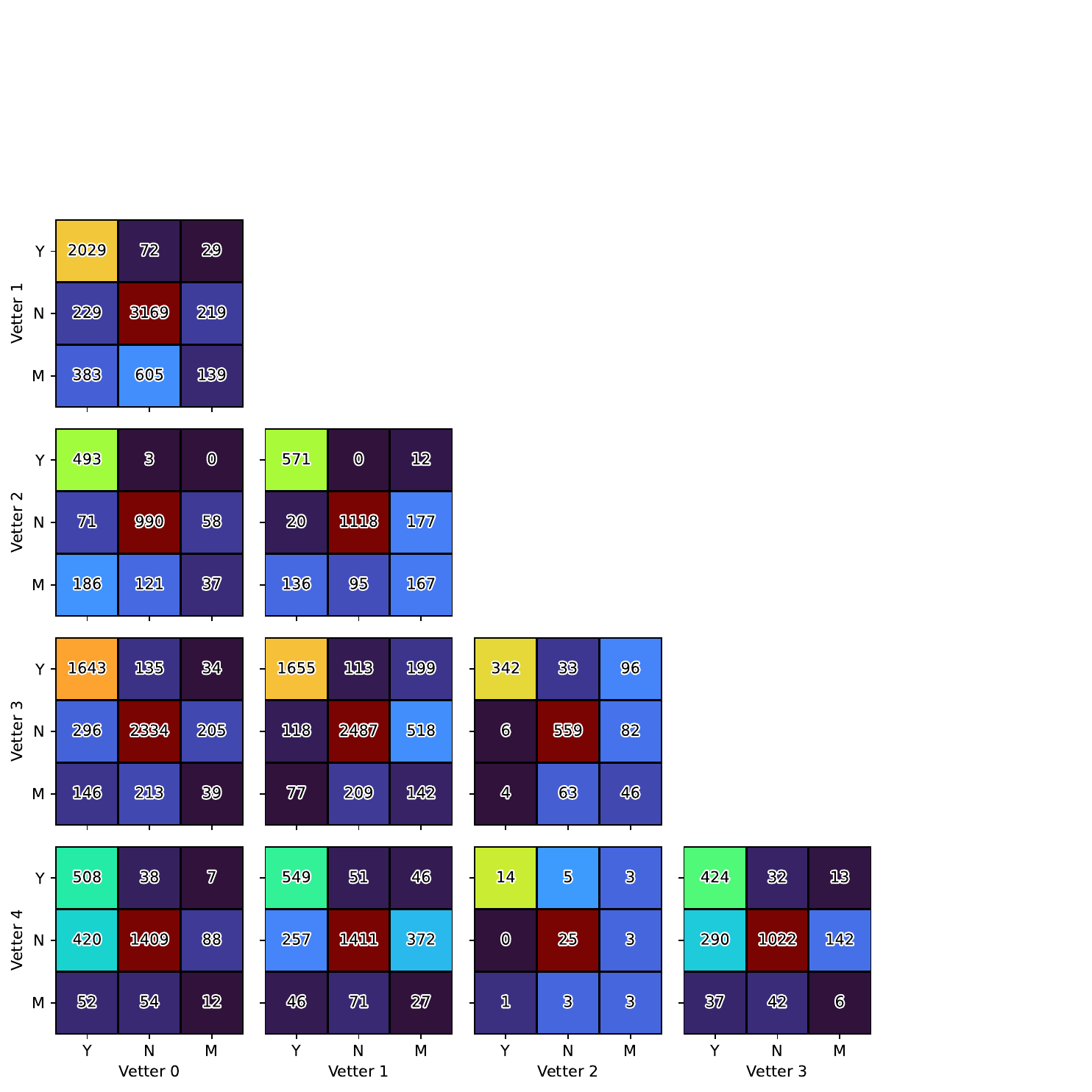}
    \caption{Correlations between vetters' votes on detections from the reverse stack (top left), implanted sources in the forward stack (top right), and candidate sources in the forward stack (bottom). The numbers in each square tabulate the number of times the square's outcome occurred. Note that \texttt{yes} votes on sources from the reverse stack are not strongly correlated.}
    \label{fig:voter-correlations}
\end{figure}

In principle, $P(\text{vote}|+)$, $P(\text{vote}|-)$, and $O(+)$ can all vary with magnitude and rate of motion. In practice, however, we found that $P(\text{vote}|+)$ and $P(\text{vote}|-)$ did not vary much over the range of brightness and rate that we considered for our fits in Sections \ref{sec:luminosity} and \ref{sec:H}, so we considered them to be constants. Similarly, parameterizing $O(+)$ had little effect on our detections' weights after human inspection, so we chose to treat it as constant.\footnote{We find that our true positive rate falls off for fainter sources, but since our fits neglect sources fainter than 50\% of the maximum efficiency, the true and false positive rates are fairly constant over the full range of brightness. The prior odds also fall off for fainter sources, but since our detections are dominated by faint sources, The estimation of a uniform prior only results in a slight underestimate the odds of bright sources, which end up having weights close to 1 anyway.} See Table \ref{tab:bayes} for a tabulation of our vetters' Bayes factors, and see Figure \ref{fig:voter-correlations} for correlations between vetters' votes. 

\begin{table}[H]
\renewcommand{\arraystretch}{1.0}
\centering
\begin{tabular*}{0.4\textwidth}{l|@{\extracolsep{\fill}}rcc}
\hline
\hline
Person & B(\texttt{yes}) & B(\texttt{no}) & B(\texttt{maybe}) \\ 
\hline
    Vetter 0 & 17.50 & 0.12 & 0.49 \\ 
    \hline
    Vetter 1 & 76.96 & 0.10 & 1.40 \\
    \hline
    Vetter 2 & 849.50 & 0.10 & 3.17 \\ 
    \hline
    Vetter 3 & 42.03 & 0.19 & 1.09 \\ 
    \hline
    Vetter 4 & 23.54 & 0.35 & 1.76 \\ 
    \hline
\end{tabular*}
\caption{Anonymized Bayes factors of each of our vetters.}
\label{tab:bayes}
\end{table}

\section{Detections}
\label{sec:detections}
\begin{figure}[h]
    \centering
    \includegraphics[width=0.7\textwidth]{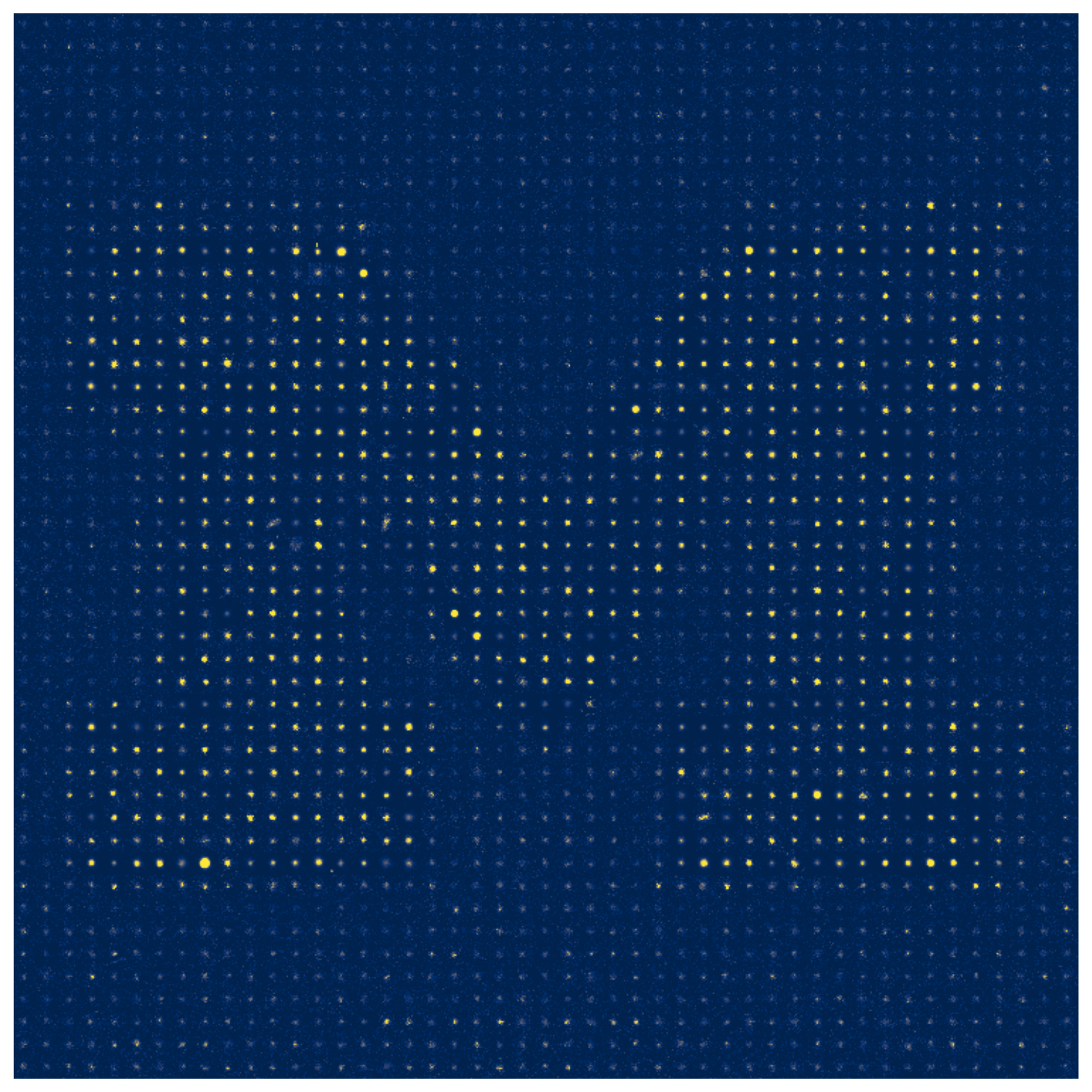}
    \caption{Mosaic of DEEP B1 detections with weight $\geq 0.4$.}
    \label{fig:mosaic}
\end{figure}

In this section we qualitatively analyze our detections; we do more thorough quantitative analyses in Sections \ref{sec:efficiency}--\ref{sec:H}. In our $20$ nights of data we detected a weighted sum of $2297.9$ objects with weight greater than $0.01$, corresponding to $2896$ unique sources. We have elected to omit $3698$ sources with weight less than $0.01$, as such detections are rather unlikely to be real, and their omission does not change the results of our analysis. While the majority of our remaining sources have weight close to 1, there are some more ambiguous cases. We show a mosaic of all detections with weight $\geq 0.4$ in Figure \ref{fig:mosaic}.

\begin{figure}[h]
    \centering
    \includegraphics[width=0.9\textwidth]{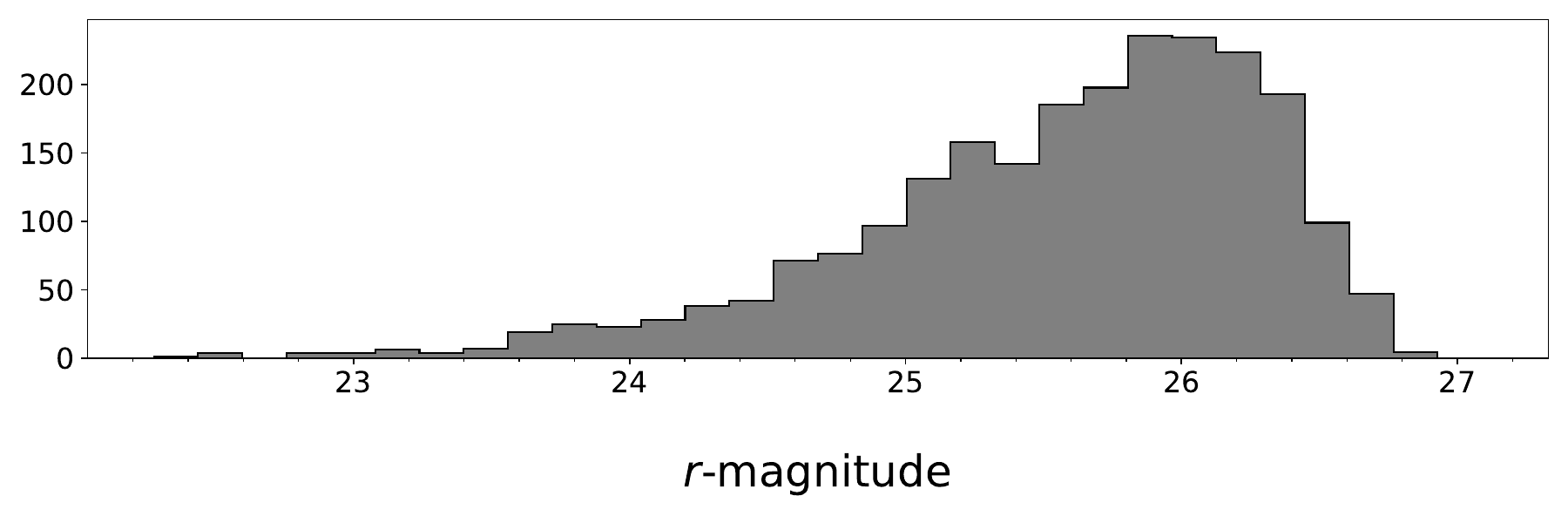}
    \caption{Weighted distribution of the apparent magnitudes of our detections (where the weights are given by Equation [\ref{eq:bayes-probability}]).}
    \label{fig:magnitudes}
\end{figure}

Based on our distribution of observed magnitudes (shown in Figure \ref{fig:magnitudes}), it appears that our detection efficiency begins to fall off at magnitudes fainter than $r \sim 26$ (though we explore this in detail in Section \ref{sec:efficiency}).

\begin{figure}[h]
    \centering
    \includegraphics[width=0.9\textwidth]{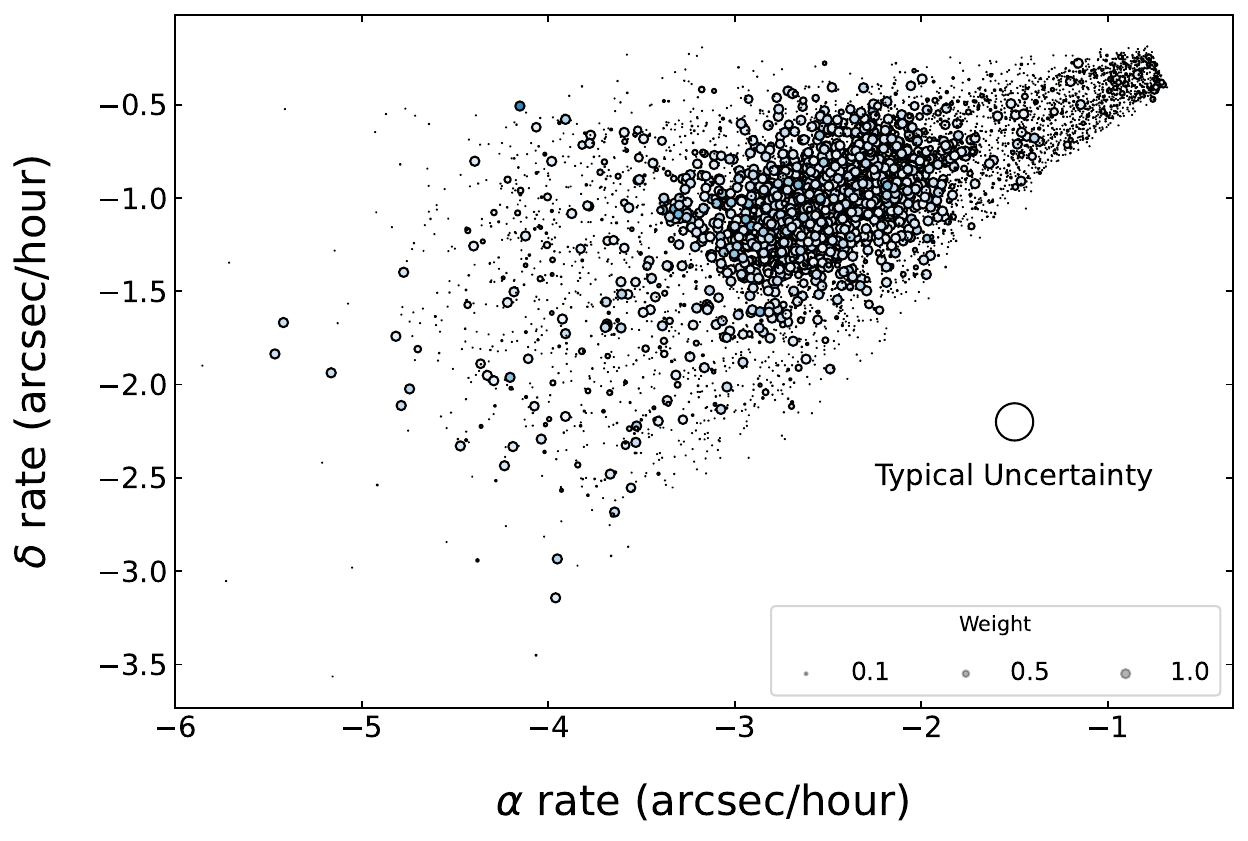}
    \caption{Sky moving rates of our candidate detections. The size of each marker is proportional to its weight, as calculated by Equation (\ref{eq:bayes-probability}). The dense cloud of points corresponds mostly to CC detections (see Section \ref{sec:dynamics}).}
    \label{fig:detection-rates}
\end{figure}

It is also informative to examine the sky moving rates of our detections, which we display in Figure \ref{fig:detection-rates}. In this figure, the size of each marker is proportional to its weight. The most apparent feature here is a large population of objects moving at approximately 3'' per hour, mostly corresponding to CCs (see Section \ref{sec:dynamics}). 
Based on the increased density of small points at slow rates of motion, we note a propensity for slow-moving false positives to pass our CNNs, but to be later given low weights after human inspection (there is a noticeable overdensity of points at slow rates in Figure \ref{fig:rates} which is absent in Figure \ref{fig:detection-rates}). At faster rates, points with low weights appear to be evenly spread. The presence of such features provides further evidence that a great deal of care in avoiding false positives is required when making statistical use of single-night detections. When such detections can be linked to several epochs, single-night false positives will be less problematic.

\section{Detection Efficiency}
\label{sec:efficiency}

\begin{figure}[h]
    \centering
    \includegraphics[width=0.95\textwidth]{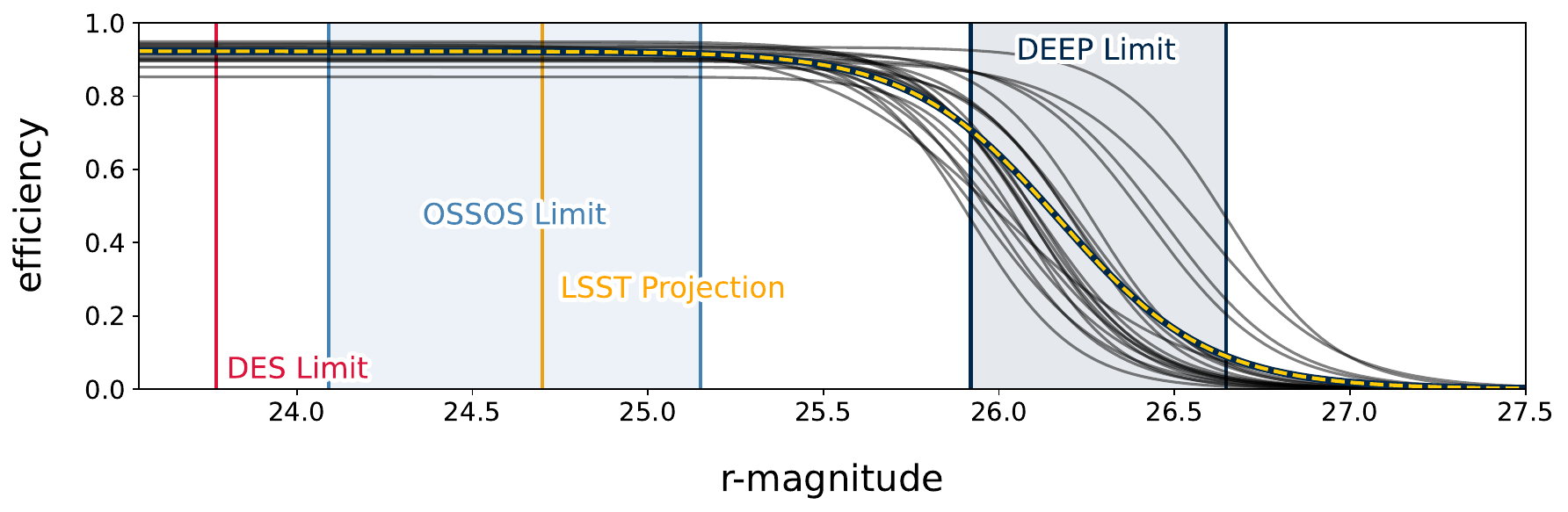}
    \caption{Recovery efficiency for implanted sources as a function of r-band magnitude. The combined efficiency for all $20$ nights of data is given by the dashed line, while individual nights’ efficiencies are given by the grey lines. The average $m_{50}$ for our entire survey is $m_r \sim 26.2$, with individual nights ranging from 25.92--26.65. Our average peak efficiency, $\eta_0$, is $\gtrsim 0.92$, with individual nights ranging from 0.85--0.95. For reference, we also show limits from the Dark Energy Survey (DES) \citep{bernardinelli2022}, OSSOS \citep{OSSOSVII} and LSST \citep{LSST}.}
    \label{fig:efficiency}
\end{figure}

In order to make use of our detections, we must understand the efficiency with which we recover our implanted synthetic sources. We parameterize the detection efficiency as a function of apparent magnitude using a single hyperbolic tangent function, given the following equation
\begin{equation}
     \eta(m) = \frac{\eta_0}{2} \left(1 - \tanh\left(\frac{m - m_{50}}{\sigma}\right) \right),
     \label{eq:efficiency}
\end{equation}
where $\eta_0$ is the peak detection efficiency, $m_{50}$ is the magnitude at which the detection efficiency drops to $\eta_0 / 2$, and $\sigma$ is the width of the hyperbolic tangent function.\footnote{More complicated functional forms do not improve the fit, and do not change the results of our analysis.} We weight each of the detections in our fit using Equation (\ref{eq:bayes-probability}), and maximize the likelihood given by
\begin{equation}
    \ln{\mathcal{L(\theta)}} = \sum_i w_i \ln\left(\eta(m_i|\theta)\right) + (1 - w_i) \ln\left(1 - \eta(m_i|\theta)\right),
\end{equation}
where $\theta$ is the vector of function parameters, and undetected fakes receive $w = 0$. We display the best fit for each night in Figure \ref{fig:efficiency}, and list the fit parameters in Table \ref{tab:B1_fields}. 

\section{The Luminosity Function}
\label{sec:luminosity}
We use our characterized detections to compute the differential sky density $\Sigma$ of the Kuiper Belt as a function of apparent magnitude.\footnote{This quantity is colloquially referred to as the luminosity function.} We reiterate that for this analysis we are using only single-epoch data (i.e., we have not linked these detections across multiple epochs), so we must treat each night as an independent survey. However, because our survey was designed to detect objects multiple times, our nights are not all statistically independent. As such, we selected a subset of our data consisting of the statistically independent set of nights \{B1b 20201015, B1c 20201016, B1e 20201017, B1a 20201018, B1f 20201020, B1d 20201021\} to do our fits. This subset offers the best combination of survey area and depth among all possible subsets of our data.\footnote{We also experimented with other subsets of our data, and all yielded results consistent with the results presented here.} Note that when fitting our data, we truncate our detection efficiency at $m_{50}$, and ignore all fainter detections.

For a given probability distribution $\Sigma(m)$, the expected number of detections by a survey is given by 
\begin{equation}
    \bar{N} = \Omega \int \eta(m) \Sigma(m | \theta) dm
\end{equation}
where $\Omega$ is the survey’s areal coverage and $\eta(m)$ is its detection efficiency. The variable $\theta$ is a vector of function parameters. Next, the probability of randomly drawing an object with magnitude $m$ from $\Sigma(m)$ is given by 
\begin{equation}
    P(m| \theta) = \int \Sigma(m' | \theta) \epsilon(m'|m, \delta m)\, dm'
\end{equation}
where $\epsilon$ is a functional representation of the magnitude uncertainty for which we have adopted a Gaussian centered at $m$, with a width of $\delta m$.\footnote{For completeness, we note that the distribution of uncertainties in the magnitudes is not exactly Gaussian. More specifically, we estimate that the true normalized distribution is slightly narrower than a true Gaussian, with slightly more power in the tails (where the distributions are not shown here). The difference in full-width-half-maximum between the two distributions is only about 10 percent. As a result, we use the Gaussian approximation for simplicity.}

We calculate the underlying luminosity function ($\Sigma$) of the Kuiper Belt by maximizing the likelihood ($\mathcal{L}$) given by
\begin{equation}
    \mathcal{L}(\theta) = \prod_{k=1}^{n} e^{-\bar{N}_k} \prod_{j=1}^{N_k} P(m_{j,k}| \theta)^{w_{j,k}}
    \label{eq:mag-L}
\end{equation}
(see, e.g., \citealt{Loredo2004}, \citealt{fraser2014}) where the index $k$ runs through each night of data, and the index $j$ runs through a survey's detections. The value $w_{j,k}$ denotes the weight of the $j$th object detected by the $k$th survey, as calculated by Equation (\ref{eq:bayes-probability}).\footnote{By applying a weight to each of our candidate detections, we are properly accounting for false positives in our data. This weight term, which is not present in other KBO analyses of this nature, was first derived by \citet{weighted-likelihood}.}

Several previous works have studied the form of the luminosity function for KBOs \citep{bernstein2004, Petit2006, Fraser2008, Fraser2009, Fuentes2009}. Following the example of these studies, we fit our data with functional forms of varying complexity. 
We first try a single power law given by 
\begin{equation}
\Sigma_{single}(m) = 10^{\alpha(m - m0)},
\label{eq:mag-single-pdf}
\end{equation}
where $m_0$ is the magnitude at which the density of objects is one per square degree, and $\alpha$ is the power law slope. We next try a rolling power law given by
\begin{equation}
\Sigma_{rolling}(m) = \Sigma_{23} 10^{\alpha_1(m - 23) + \alpha_2(m - 23)^2},
\label{eq:mag-rolling-pdf}
\end{equation}
where $\Sigma_{23}$ is the number of objects with $m_r = 23$ per square degree, while $\alpha_1$ and $\alpha_2$ control the shape of the function. We fit a broken power law given by
\begin{equation}
\Sigma_{broken}(m) = 
  \begin{cases}
    10^{\alpha_1(m - m_0)}  & \quad m < m_B \\
    10^{\alpha_2(m - m_0) + (\alpha_1 - \alpha_2)(m_B - m_0)} & \quad m \geq m_B
  \end{cases}
\label{eq:mag-broken-pdf}
\end{equation}
were $m_0$ is a normalization parameter, $m_B$ is the magnitude at which the break occurs, and $\alpha_1$ and $\alpha_2$ are the bright-and-faint slopes, respectively. Finally we fit an exponentially tapered power law with the functional form 
\begin{equation}
  \Sigma_{taper}(m) = \exp\left[-10^{\beta(m_B - m)}\right] 10^{(\alpha - \beta)m - (m_0 \alpha)} \left[\alpha 10^{m \beta} + \beta 10^{m_B \beta}\right]
  \label{eq:taper-pdf-mag}
\end{equation}
where $\alpha$ is the faint-end power law slope, $\beta$ is the strength of the exponential taper, $m_0$ is a normalization parameter, and $m_B$ is the magnitude at which the exponential taper begins to dominate.

\begin{figure}[h]
    \centering
    \includegraphics[width=1\textwidth]{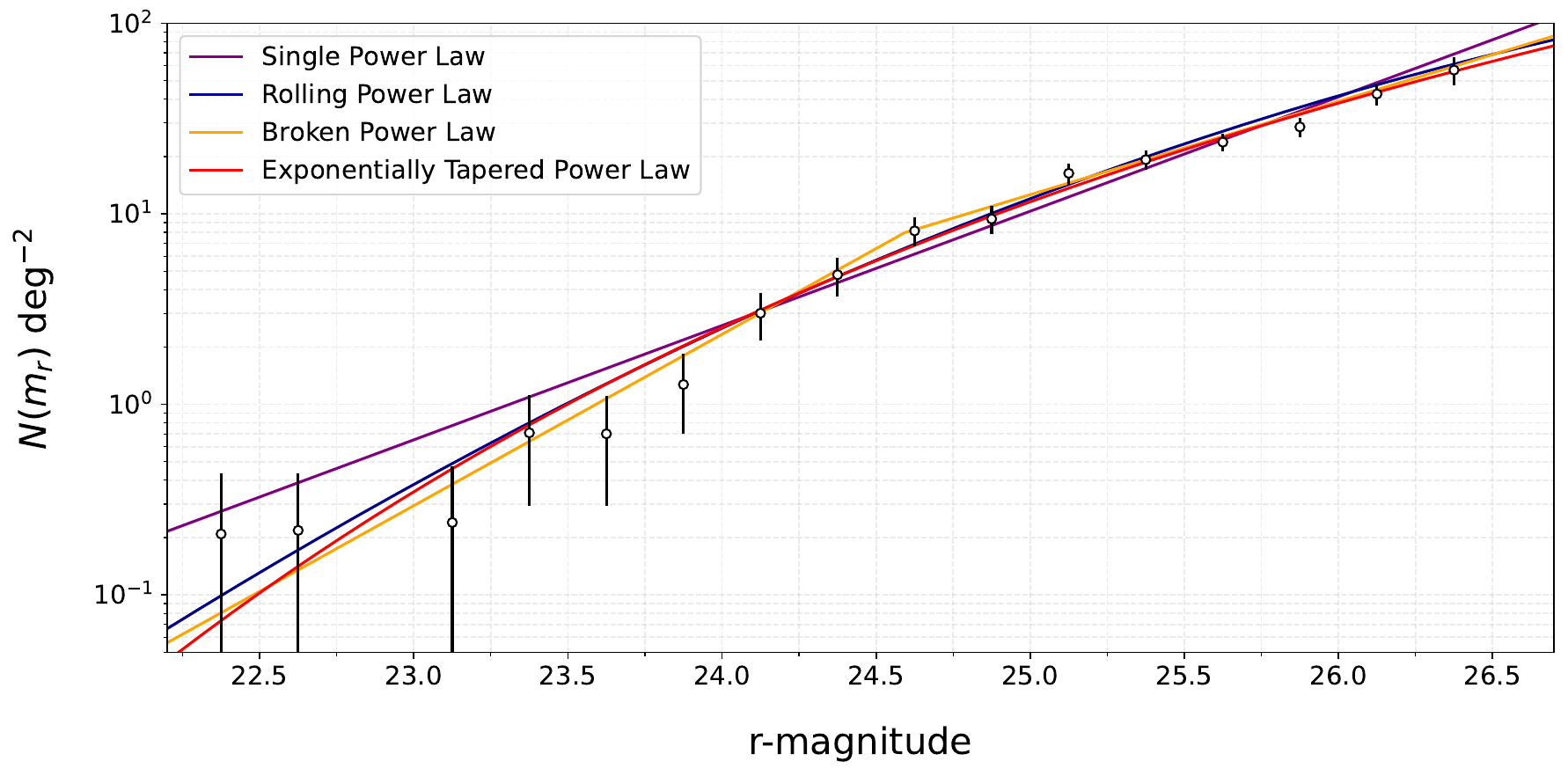}
    \caption{Best-fit differential distributions for the single, rolling, broken, and exponentially tapered power laws in purple, blue, yellow, and red respectively. The points and 1-$\sigma$ error bars represent the differential distribution of our detections, corrected for efficiency and weight. The points are meant only as a visual aid; we always use the maximum likelihood technique to fit our data. Note that our fits are not normalized with respect to latitude, but rather they represent the average sky density among the fields in our pointing mosaic.}
    \label{fig:luminosity-demonstration}
\end{figure}

In each case, we obtain a best fit using an optimizer, and then estimate our uncertainties by running an MCMC. We then use a survey simulation technique to test the quality of our fits. We randomly sample a population of objects from our best-fit $\Sigma(m|\theta)$, and then impose the detection criteria of our survey to simulate detections. After we simulate our detections, we construct a simulated empirical distribution, $E(m)$. By repeating this process many times, we end up with an ensemble of simulated empirical distribution functions, $\{E_i(m)\}$. This ensemble is representative of the distribution of possible outcomes for our survey under the assumption that our best fit $\Sigma(m|\theta)$ is the underlying truth. Next we calculate the upper and lower limits for the central 95th percentile of $\{E_i(m)\}$ as a function of $m$, which we call $u(m)$ and $l(m)$, respectively. We use these values to define a test statistic that measures the fraction of the interval $m$ over which $E_i(m)$ is an outlier at the 95 percent level. The statistic, which we call $S$, is given by
\begin{equation}
    S = \int f(m) \, dm
\end{equation}
where
\begin{equation}
    f(m) = \begin{cases}
    0, & \text{if} \quad l(m) < E(m) < u(m) \\
    1, & \text{otherwise}.
  \end{cases}
\end{equation}
We calculate $S$ for each simulated survey to assemble a set $\{S_i\}$, and then for our actual detections, $S_d$. We then compute the quantile of $S_d$ among $\{S_i\}$, which we call $Q_{\text{outlier}}$. Values of $Q_{\text{outlier}} > 0.95$ indicate a poor fit. Finally, we compute the Bayes Information Criterion (BIC) for each of our fits. In general, lower values of BIC indicate a preferred model. We demonstrate our results in Figure \ref{fig:luminosity-demonstration}, and summarize them in Table \ref{tab:luminosity}.

\begin{table}[h]
\renewcommand{\arraystretch}{1.25}
\centering
\begin{tabular*}{1\textwidth}{l@{\extracolsep{\fill}}lccc}
\hline
\hline
Functional Form & Best-Fit Parameters & BIC & $Q_{\text{outlier}}$\\ \hline
    Single Power Law & $\langle \alpha, m_0\rangle = \langle 0.60^{+0.02}_{-0.02}, 23.31^{+0.09}_{-0.10} \rangle$ & $5.2$ & $0.99$ \\ 
    \hline
    Rolling Power Law & $\langle \alpha_1, \alpha_2, \Sigma_{23}\rangle = \langle 0.89^{+0.11}_{-0.10}, -0.07^{+0.02}_{-0.03}, 0.38^{+0.11}_{-0.09} \rangle$ & $0$ & $0.19$ \\ 
    \hline
    Broken Power Law & $\langle \alpha_1, \alpha_2, m_0, m_B\rangle = \langle 0.90^{+0.16}_{-0.12}, 0.49^{+0.04}_{-0.05}, 23.59^{+0.09}_{-0.1a}, 24.59^{+0.26}_{-0.37} \rangle$ & $2.1$ & $0.18$ \\
    \hline
    Tapered Power Law & $\langle \alpha, \beta, m_0, m_B\rangle = \langle 0.21^{+0.10}_{-0.08}, 0.14^{+0.05}_{-0.04}, 13.95^{+5.33}_{-11.40}, 28.74^{+2.71}_{-2.18} \rangle$ & $7.9$ & $0.83$ \\
\hline
\end{tabular*}
\caption{Best fit parameters and statistics for each of the distributions we tested on the full KBO sample. The BICs have been normalized such that the minimum value among the distributions is 0.}
\label{tab:luminosity}
\end{table}

First we note that like several studies before us, we strongly rule out the single power law. The rolling, broken, and exponentially tapered power laws all provide acceptable fits. Judging by the BIC, the rolling power law is marginally preferable to the broken and exponentially tapered power laws, but we contend that a dearth of detections brightward of $m_r = 24$ limits the usefulness of such comparisons. There is also no overwhelming physical motivation for choosing between the distributions. While \citet{Kavelaars2021} showed that the absolute magnitude distribution of the CCs is well-described by an exponentially tapered power law, there is no reason \textit{a priori} that it should fit the full Kuiper Belt luminosity function. In fact, we might expect the full Kuiper Belt luminosity function to be fit poorly by an exponentially tapered power law, because it is likely a mix of multiple distributions. However, as we show in Section \ref{sec:dynamics}, our detections are dominated by CCs. It is therefore unsurprising that the exponentially tapered power law yields a reasonable fit. Given these considerations, we opt not to choose a preferred model, and instead claim for the time being that all three distributions provide acceptable fits to the DEEP detections.

\section{Isolating a Sample of Cold Classicals}
\label{sec:dynamics}
\begin{figure}[h]
    \centering
    \includegraphics[width=0.9\textwidth]{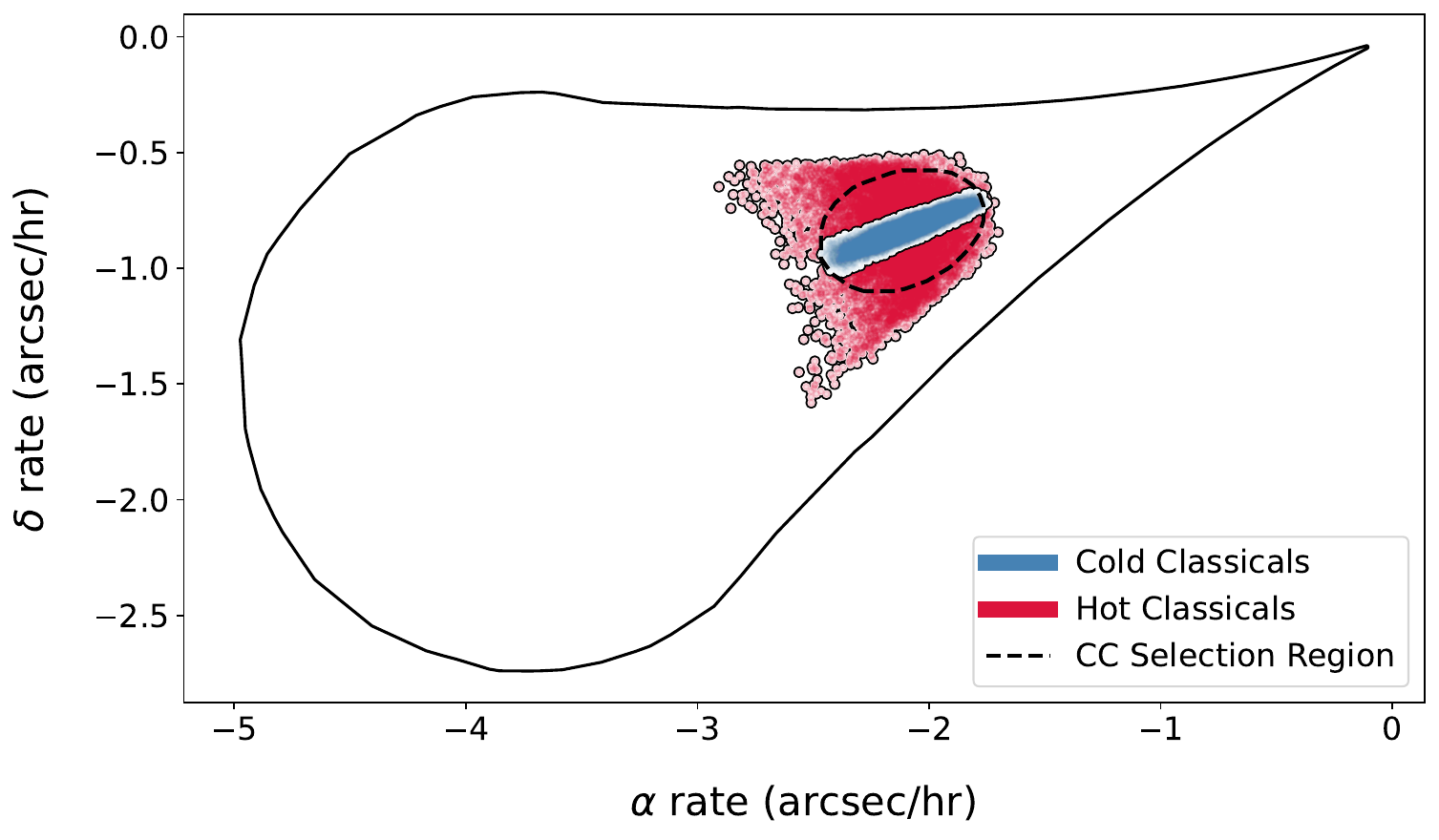}
    \caption{Sky motion parameter space for simulated CCs and HCs in B1a on 2020-10-18. The dashed line shows the CC selection region for this night.}
    \label{fig:dynamics}
\end{figure}

While our single-night detections do not provide nearly enough information for secure dynamical classification, we can use the detections' rates of motion to make assumptions about their dynamical classes. Consider the space of allowable sky motions for bound orbits shown in Figure \ref{fig:dynamics}. As before, the large black outline contains the space of possible rates of motion for objects on bound orbits at topocentric distances of 35--1000 au. The blue region is the allowable parameter space for CCs ($42.4\text{ au} < a < 47.7\text{ au}$, $0 < e < 0.2$, $0\degree < i < 4\degree$), and the red region is the allowable parameter space for a somewhat arbitrary definition of Hot Classicals (HCs) that we are only using for the purpose of demonstration ($42.4 \text{ au} < a < 47.7 \text{ au}$, $0 < e < 0.2$, $4\degree < i < 45\degree$). Note that the exact shape and orientation of these regions varies as a function of sky position and epoch.

In practice, there is some overlap in the rates of motion between different dynamical classes, which is accentuated by the uncertainty in our rate measurements. We use survey simulation to isolate the CCs in our detections. Using the OSSOS\texttt{++} Kuiper Belt model\footnote{The OSSOS++ model is designed to be a realistic model of the Kuiper Belt, including correct relative population sizes of the different Kuiper Belt dynamical classes \citep{Kavelaars2021, Crompvoets2022}.}, we calculate which objects our survey would have detected, and apply a smear in the RA and Dec rates consistent with our real detections. For each simulation we use a kernel density estimator to draw a region containing 95\% of CCs. We find that we can isolate a fairly pure sample (on average $\sim 70$\% purity; see Table \ref{tab:detections}) of CCs from our single-night data. This purity is somewhat serendipitous, as the DEEP B1 fields happen to be in a patch of sky that is relatively uncontaminated by objects in resonance with Neptune. We speculate that the uncertainties inherent to the distribution of objects in the OSSOS\texttt{++} model are larger than the purely statistical uncertainties listed in Table \ref{tab:detections}. Nevertheless, these values should provide a realistic estimate of the purity of the CC samples that we attempt to isolate for the analysis in Section \ref{sec:H}.

\begin{table}[h]
\renewcommand{\arraystretch}{0.95}
\centering
\begin{tabular*}{0.93\textwidth}{lc|rr|ccc|c}
\hline
\hline
    Night & Field & Total Detections & Potential CCs & CC Fraction & HC Fraction & Resonant Fraction & CC fits\\ 
    \hline
    2019-08-27 & B1c & 89.94  & 69.91  & 0.68 $\pm$ 0.06 & 0.19 $\pm$ 0.04 & 0.12 $\pm$ 0.05 & no \\
    2019-08-28 & B1a & 78.39  & 64.56  & 0.74 $\pm$ 0.05 & 0.15 $\pm$ 0.03 & 0.11 $\pm$ 0.03 & no \\
    2019-08-29 & B1b & 160.83 & 131.18 & 0.74 $\pm$ 0.04 & 0.15 $\pm$ 0.03 & 0.10 $\pm$ 0.02 & no \\
    \hline
    2019-09-26 & B1a & 81.58  & 65.10  & 0.75 $\pm$ 0.05 & 0.14 $\pm$ 0.04 & 0.11 $\pm$ 0.04 & no \\
    2019-09-27 & B1b & 77.85  & 56.96  & 0.74 $\pm$ 0.06 & 0.16 $\pm$ 0.05 & 0.09 $\pm$ 0.03 & no \\
    2019-09-28 & B1c & 118.20 & 88.98  & 0.68 $\pm$ 0.06 & 0.18 $\pm$ 0.04 & 0.13 $\pm$ 0.04 & no \\
    \hline
    2020-10-15 & B1b & 68.72  & 53.86  & 0.71 $\pm$ 0.04 & 0.18 $\pm$ 0.04 & 0.10 $\pm$ 0.03 & yes \\
    2020-10-16 & B1c & 106.54 & 85.13  & 0.66 $\pm$ 0.05 & 0.19 $\pm$ 0.05 & 0.13 $\pm$ 0.04 & yes \\
    2020-10-17 & B1e & 120.69 & 103.62 & 0.75 $\pm$ 0.05 & 0.15 $\pm$ 0.04 & 0.09 $\pm$ 0.03 & yes \\
    2020-10-18 & B1a & 124.90 & 102.37 & 0.71 $\pm$ 0.04 & 0.17 $\pm$ 0.04 & 0.11 $\pm$ 0.02 & yes \\
    2020-10-19 & B1d & 54.50  & 42.09  & 0.66 $\pm$ 0.07 & 0.21 $\pm$ 0.05 & 0.12 $\pm$ 0.04 & no \\
    2020-10-20 & B1f & 56.72  & 38.71  & 0.57 $\pm$ 0.08 & 0.23 $\pm$ 0.05 & 0.18 $\pm$ 0.06 & no \\
    2020-10-21 & B1d & 76.85  & 53.24  & 0.66 $\pm$ 0.06 & 0.21 $\pm$ 0.05 & 0.11 $\pm$ 0.04 & yes \\
    \hline
    2021-09-27 & B1d & 68.32  & 58.31  & 0.68 $\pm$ 0.06 & 0.19 $\pm$ 0.05 & 0.12 $\pm$ 0.04 & no \\
    2021-10-01 & B1b & 110.33 & 85.49  & 0.71 $\pm$ 0.05 & 0.18 $\pm$ 0.05 & 0.10 $\pm$ 0.02 & no \\
    2021-10-02 & B1f & 80.02  & 56.98  & 0.55 $\pm$ 0.07 & 0.26 $\pm$ 0.07 & 0.18 $\pm$ 0.06 & no \\
    2021-10-03 & B1i & 114.10 & 92.96  & 0.69 $\pm$ 0.05 & 0.18 $\pm$ 0.03 & 0.12 $\pm$ 0.04 & no \\
    2021-10-04 & B1c & 84.95  & 61.39  & 0.68 $\pm$ 0.06 & 0.18 $\pm$ 0.03 & 0.13 $\pm$ 0.05 & no \\
    2021-10-05 & B1h & 109.67 & 90.02  & 0.74 $\pm$ 0.04 & 0.15 $\pm$ 0.04 & 0.10 $\pm$ 0.03 & no \\
    2021-10-06 & B1e & 103.83 & 81.73  & 0.73 $\pm$ 0.06 & 0.16 $\pm$ 0.05 & 0.10 $\pm$ 0.04 & no \\
    \hline
\end{tabular*}
\caption{(Second column) Weighted number of detections in each of our 20 fields, along with the weighted number of objects consistent with being CCs. Note that we have ignored all objects fainter than the $m_{50}$ of the night in which they were detected. (Third column) Fraction by dynamical class of the objects with rates consistent with being CCs, as determined by survey simulation. The final column indicates whether a field was used to derive constraints for the CC population.}
\label{tab:detections}
\end{table}

Finally, we obtain a distance estimate and uncertainty for each of our CC detections. This estimate relies on the fact that in the regime of the CCs, the relationship between an object's heliocentric distance and the inverse of its apparent rate of motion can be well-approximated as linear. For each night of data, we project our CC population model (obtained from the OSSOS\texttt{++} model) into the space of RA rate and Dec rate, and fit a line relating distance to the inverse of the apparent sky motion. We then use the resulting relationship to compute the heliocentric distance of each of our detections. By sampling from the covariance of our detections' rates, we obtain Gaussian distance uncertainties of at most 1--2 au, which end up being precise enough to fit an absolute magnitude distribution (Section \ref{sec:H}).

\section{The Absolute Magnitude Distribution of the Cold Classicals}
\label{sec:H}
Determining the absolute magnitude ($H$) of an object requires simultaneous knowledge of its apparent magnitude and its heliocentric distance. While our detections have reasonably well-constrained apparent magnitudes ($\sim \pm 0.1$), their distances are not as well-constrained from single-night detections \textit{a priori}. However, as we found in Section \ref{sec:dynamics}, any true CCs are within a narrow range of $r$, with uncertainties of only 1--2 au, leading to uncertainty of only 0.1--0.2 mag in $H$.

For a given probability distribution $\Sigma(H)$, the expected number of detections by a survey is given by 
\begin{equation}
    \bar{N} = \Omega \iint \eta(m)  \Sigma(H(r, m) | \theta) \Gamma(r) \, dm dr,
\end{equation}
where $\Omega$ is the survey’s areal coverage and $\Gamma$ is the underlying radial distribution of objects in the survey’s field of view. Note that we use the OSSOS\texttt{++} Kuiper Belt model \citep{Kavelaars2021} to compute a kernel density for $\Gamma$. Since our fields are all observed near opposition, it is a good approximation to use 
\begin{equation}
    H = m - 5 \log_{10}(r(r-1)). 
\end{equation}
Next, the probability of randomly drawing an object with magnitude $m$ and heliocentric distance $r$ from the distribution $\Sigma(H)$ is given by 
\begin{equation}
    P(m, r | \theta) = \iint \Sigma(H(r', m') | \theta) \epsilon(m'm'|m, \delta m)\Gamma(r')\gamma(r'|r,\delta r)\, dm'dr'
\end{equation}
where $\epsilon$ is a functional representation of the magnitude uncertainty, and $\gamma$ is a functional representation of the uncertainty in heliocentric distance. Note that we have taken $\epsilon$ to be a Gaussian centered at $m$, with a width of $\delta m$, and $\gamma$ to be a Gaussian centered at $r$, with a width of $\delta r$.

We again use a modified version of the method described by \cite{fraser2014} in which we maximize the likelihood given by
\begin{equation}
    {\mathcal L}(\theta) = \prod_{k=1}^{n} e^{-\bar{N}_k} \prod_{j=1}^{N_k} P(m_{j,k}, r_{j,k}| \theta)^{w_{j,k}}
\end{equation}
where the index $k$ runs through the surveys (in our case individual nights of data), and the index $j$ runs through a survey's detections.

We again fit single, rolling, broken, and exponentially tapered power laws. Note, however, that we have changed the definition of the rolling power law to
\begin{equation}
\Sigma_{rolling}(H) = \Sigma_{8} 10^{\alpha_1(H - 8) + \alpha_2(H - 8)^2}
\end{equation}
where $\Sigma_{8}$ is the number of objects per square degree per magnitude at $H_r=8$.

\begin{figure}[h]
    \centering
    \includegraphics[width=0.9\textwidth]{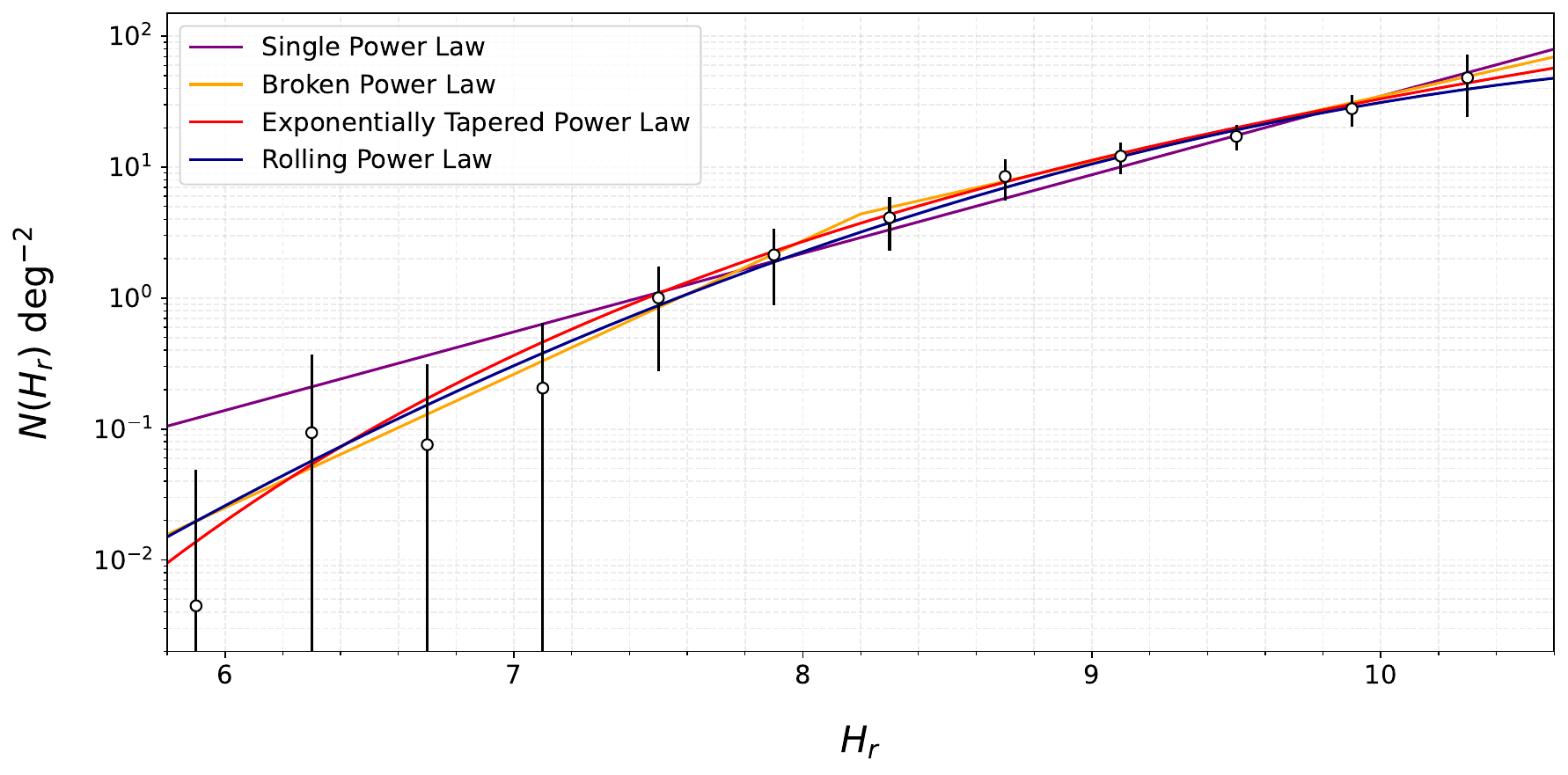}
    \caption{Best-fit differential distributions for the single, rolling, broken, and exponentially tapered power laws in purple, blue, yellow, and red, respectively. The points and 1-$\sigma$ error bars represent the differential distribution of our detections, corrected for efficiency and weight. Note that our fits are not normalized with respect to latitude, but rather they represent the average sky density among the fields in our pointing mosaic.}
    \label{fig:H-taper}
\end{figure}

For these fits, we must also account for the contamination in our sample from non-CCs. We do this by sampling from each night's detections that have rates consistent with being a CC, accepting each detection with a probability given by the CC fraction column in Table \ref{tab:detections}. Note that we omit the B1f field due to the projected low purity of the isolated CC sample. We then do the maximum likelihood fit, and compute our uncertainties as in the previous sections. We repeat this procedure 100 times, and then aggregate all of our all samples from all MCMCs, resulting in a combined posterior. We then calculate the best fit parameters and uncertainties from the aggregation. For the fitting statistics, we now report the mean of BIC and $Q_{\text{outlier}}$ from our simulations as $\langle \text{BIC} \rangle$ and $\langle Q_{\text{outlier}} \rangle$ respectively. We show our fits in Figure \ref{fig:H-taper}, and present our results in tabular form in Table \ref{tab:H}.

\begin{table}[h]
\renewcommand{\arraystretch}{1.25}
\centering
\begin{tabular*}{1\textwidth}{l@{\extracolsep{\fill}}lccc}
\hline
\hline
Functional Form & Best-Fit Parameters & $\langle \text{BIC} \rangle$ & $\langle Q_{\text{outlier}} \rangle$\\ \hline
    Single Power Law & $\langle \alpha, H_0\rangle = \langle 0.60^{+0.04}_{-0.04}, 7.43^{+0.12}_{-0.13} \rangle$ & $4.0$ & $0.86$ \\ 
    \hline
    Rolling Power Law & $\langle \alpha_1, \alpha_2, \Sigma_{8}\rangle = \langle 0.77^{+0.10}_{-0.08}, -0.10^{+0.04}_{-0.05}, 2.26^{+0.35}_{-0.33} \rangle$ & $0$ & $0.20$ \\ 
    \hline
    Broken Power Law & $\langle \alpha_1, \alpha_2, H_0, H_B\rangle = \langle 1.02^{+0.48}_{-0.23}, 0.50^{+0.07}_{-0.09}, 7.57^{+0.14}_{-0.19}, 8.20^{+0.51}_{-0.67} \rangle$ & $5.2$ & $0.66$ \\
    \hline
    Tapered Power Law & $\langle \alpha, \beta, H_0, H_B\rangle = \langle 0.26^{+0.08}_{-0.07}, 0.19^{+0.08}_{-0.06}, 2.22^{+2.42}_{-3.72}, 10.62^{+1.86}_{-1.14} \rangle$ & $5.9$ & $0.56$ \\
\hline
\end{tabular*}
\caption{Best fit parameters and statistics for each of the distributions tested for the absolute magnitude distribution of the CC subsample of our detections. The BIC values have each been rescaled such that the minimum value among the distributions is 0.}
\label{tab:H}
\end{table}

\begin{figure}[h]
    \centering
    \includegraphics[width=0.9\textwidth]{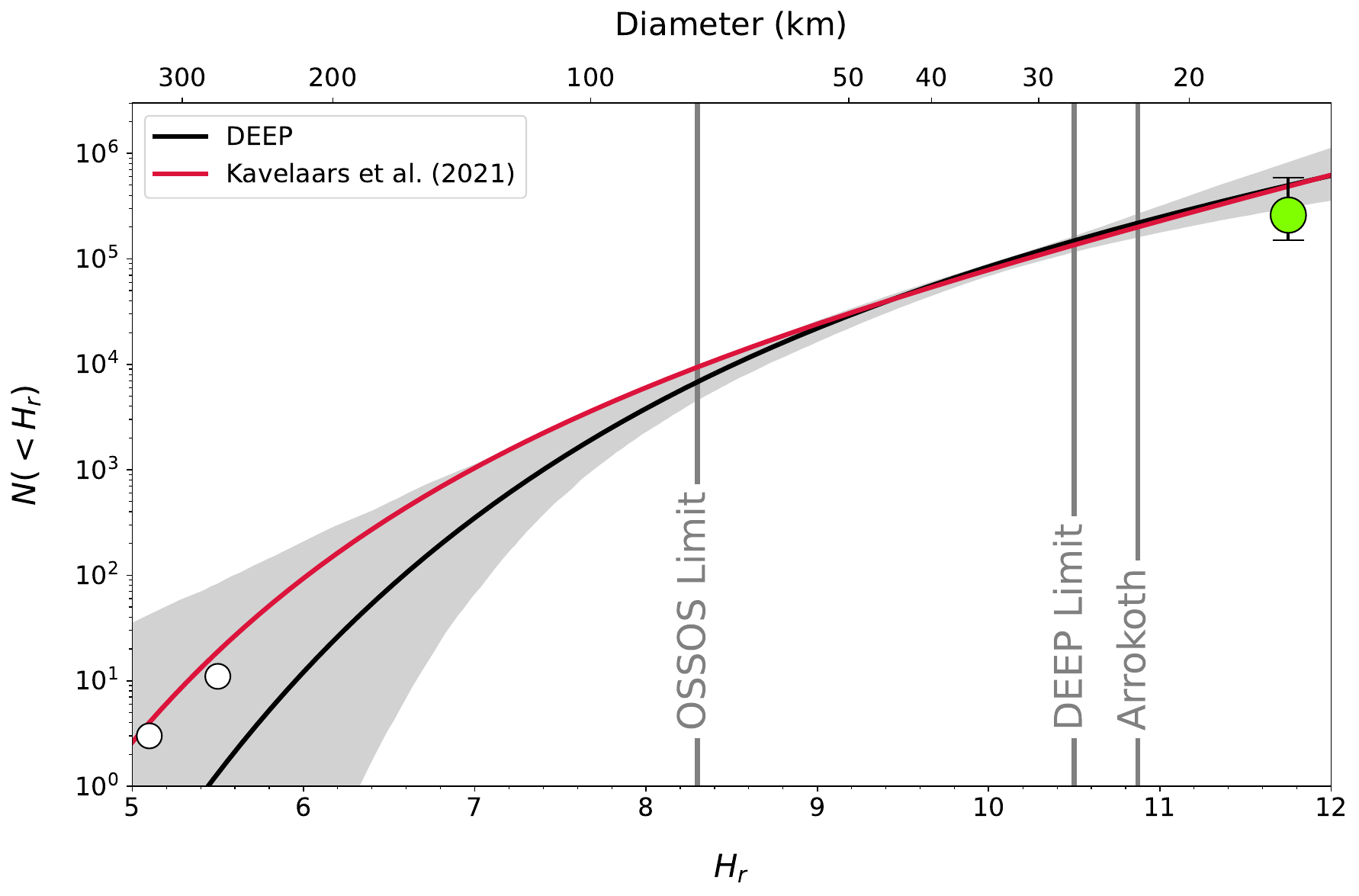}
    \caption{Absolute magnitude distribution of the CCs in our sample compared to those in \citet{Kavelaars2021}. The black line is our best fit exponentially tapered power law, and the dark grey region represents a 95\% confidence interval. Note that we have no detections brighter than $H_r \sim 6.3$. The red line is the best fit from \citet{Kavelaars2021}, and the white and green circles were taken from the same work. The white circles represent where the inventory of the CCs is considered complete, and the green circle is computed using the detections from \citet{bernstein2004}.}
    \label{fig:H-comparison}
\end{figure}

While the BIC favors the rolling power law, we don't believe that we can give preference to any distribution. This lack of constraining power is largely due to our dearth of detections at the bright end of the $H$ distribution (we have no CC detections brighter than $H_r \sim 6.3$). In contrast, \citealt{Kavelaars2021} found that the OSSOS\texttt{++} detections were inconsistent with rolling and broken power laws due to a taper at the bright end of the $H$ distribution. The OSSOS\texttt{++} model sensitivity was only possible due to the high purity of their sample and its well-characterized orbits. Our inability to rule out the rolling and broken power laws may be due to some combination of a relatively small survey area and contamination in our CC sample. On the other hand, the OSSOS\texttt{++} sample only went as deep as $H_r \sim 8.3$ (at least two magnitudes brighter than DEEP), requiring the assumption of a faint-end power law. In contrast, we constrain the faint end of the $H$ distribution rather well. Our measured faint-end slope, $\alpha = 0.26^{+0.08}_{-0.07}$, is consistent with SI simulations \citep{Abod2019}. In Figure \ref{fig:H-comparison} we show a comparison of our exponentially tapered power law fit with that of \citet{Kavelaars2021}. The two results are consistent (i.e., the OSSOS result is within our 95\% confidence region), and both surveys are consistent with the result of \citet{bernstein2004}, which is the deepest KBO survey to date.

The consistency of our detections with a rolling power law warrants careful consideration. First, we note that for appropriate parameters, a rolling power law is functionally equivalent to a Gaussian. All rolling power law fits we have presented in this work satisfy these criteria, and can therefore be equally well-represented as normal distributions. If the CCs are normally distributed in $H$, it would imply that they have a characteristic size, after which the $H$ distribution turns over. While we currently have no reason to suspect that the size distribution turns over, our data cannot rule it out as a possibility. Furthermore, the only survey in the literature with data beyond our limit \citep{bernstein2004} cannot distinguish between the models, as it is consistent with extrapolations of both the rolling and exponentially tapered power laws. While more precise measurements of the CC $H$ distribution from forthcoming DEEP data will help to bring the true distribution into clearer focus, an even deeper targeted CC survey may be necessary to distinguish between models.

\section{Mass of the Cold Classical Kuiper Belt}
\label{sec:mass}
Given our absolute magnitude distribution, we can calculate the total mass of the CCs as
\begin{equation}
    M_{CC} = \frac{1}{F}\int \Sigma(H|\theta) M(H|p,\rho) \, dH
    \label{eq:mass}
\end{equation}
where $F$ is an estimate of the average fraction of the total CC population per sq.~deg.~in one of our fields (we calculate $F \approx 3.27 \times 10^{-4}$ using the OSSOS\texttt{++} model), and $M$ is the mass of a body as a function of $H$ given a geometric albedo, $p$, and a mass density, $\rho$. If we assume that $\rho$ and $p$ are constant among the population of CCs, we can manipulate Equation (\ref{eq:mass}) to pull the $\rho$ and $p$ dependence out of the integral, as
\begin{equation}
    M_{CC} = \frac{\rho}{F} \frac{1329^3 \pi}{6 p^{3/2}}\int \Sigma(H|\theta) 10^{-0.6 H} \, dH.
    \label{eq:mass-split}
\end{equation}
Note that the quantity $$\frac{1329^3 \pi}{6 p^{3/2}}\int \Sigma(H|\theta) 10^{-0.6 H} \, dH$$ gives the volume of CCs per square degree in units of km$^3$ deg$^{-2}$.

We find that $M_{CC}(H_r < 10.5)$ (i.e., the mass of the CCs up to our detection limit) is (at 95\% CL)
\begin{equation}
    M_{CC}(H_r < 10.5) = 0.0014^{+0.0008}_{-0.0003} M_\earth 
    \times \left( \frac{p}{0.15} \right)^{-3/2} 
    \times \left( \frac{\rho}{1.0 \text{ g cm}^{-3}} \right) 
    \times \left( \frac{F}{3.27 \times 10^{-4}  \text{ deg}^{-2}} \right)^{-1}
\end{equation}
If we extrapolate our fits our to $H_r = 12$, we find
\begin{equation}
    M_{CC}(H_r < 12) = 0.0017^{+0.0010}_{-0.0004} M_\earth 
    \times \left( \frac{p}{0.15} \right)^{-3/2} 
    \times \left( \frac{\rho}{1.0 \text{ g cm}^{-3}} \right) 
    \times \left( \frac{F}{3.27 \times 10^{-4} \text{ deg}^{-2}} \right)^{-1}.
\end{equation}

To facilitate a comparison, we can modify the form of the mass estimate reported by \citet{bernstein2004}, to 
\begin{equation}
    M_{CC-B04}(H_r \lesssim 14) = 0.0016^{+0.0003}_{-0.0003} M_\earth 
    \times \left( \frac{p}{0.15} \right)^{-3/2} 
    \times \left( \frac{\rho}{1.0 \text{ g cm}^{-3}} \right) 
    \times \left( \frac{r}{43.8 \text{ au}} \right)^{6}
\end{equation}
where $r$ is the average heliocentric distance of a CC.\footnote{\citet{bernstein2004} measured a luminosity function rather than an $H$ distribution, and thus had to estimate $r \approx 42$. Recent CC detections from both OSSOS \citep{OSSOSVIII} and DES \citep{bernardinelli2022} have shown that a more appropriate approximation is $r \approx 43.8$ au.} While the mass estimate from \citet{bernstein2004} goes slightly deeper than our extrapolation, the extra mass is negligible, so the two mass estimates are in excellent agreement. Finally, we note that since our $H$ distribution is consistent with that found by \citet{Kavelaars2021}, our mass estimates must also be in agreement.

\section{Consistency with Deeper surveys}
\label{sec:b04}
The only survey in the literature that is significantly deeper than the present work is that of \citet{bernstein2004}, which reached $m_{50} = 29.02$\footnote{We are using the correction $r - m_{F606W} = 0.15$.} over a search area of 0.019 $\text{deg}^2$. Although the survey area is quite small, we can use it as a powerful lever arm to determine whether our fits remain valid down to $H_r \sim 12$. To do so, we simulate the survey of \citet{bernstein2004}, using our $\Sigma(H)$ fits (and their uncertainties) true underlying CC $H$ distribution. For each $H$ form, we simulate the survey $10^3$ times, and then ask whether the true number of detections by the survey ($N = 3$) is commensurate with the suite of simulations. In particular, we calculate $P(\leq N)$, the probability that the survey would have made fewer than or exactly 3 detections. For the broken power law we find $P(<=3) < 0.02$, for the exponentially tapered power law we find $P(<=3) = 0.16$, and for the rolling power law we find $P(<=3) = 0.65$.

First, we note that our best-fit exponentially tapered and rolling power laws are both consistent with the B04 detections. We are hesitant to rule out the broken power law because we find that dropping the $m_{50}$ of B04 from $r \sim 29.02$ to $r \sim 28.82$ yields $P(<=3) = 0.05$ for the broken power law. While this result is still indicative of a marginal fit, it is not strong enough to rule out the broken power law altogether. Although the B04 survey had sensitivity as faint as $r \sim 29.02$, its faintest detection was $r \sim 28.23$, so it is possible that the reported $m_{50}$ was overestimated. If, on the other hand, the reported efficiency of B04 is accurate, it is possible that the $H$ distribution of the CCs flattens out (and possibly turns over) somewhere between the limits of DEEP and B04. While DEEP will not be able to resolve this issue, a very deep targeted CC survey should be able to answer the question \citep{stansberry2021}.

\section{Discussion and Conclusions} 
\label{sec:conclude}

In this paper we have presented our single-night detections from $20$ nights of data in the DEEP B1 field. By using a shift-and-stack technique we were able to achieve an $r$-band depth of $\sim26.2$ over approximately 60 square degrees of sky. Our data yielded $2297.9$ single-epoch candidate detections, including $1849.8$ detections fainter than $m_r \sim 25$---the most detections fainter than $m_r \sim 25$ ever reported in a single survey by more than an order of magnitude.

Our claim of fractional discoveries is a first for KBO science, and as such may seem peculiar. However, as we have shown in Section \ref{sec:shiftstack}, a weighted treatment of our detections allows us to properly account for false positives. By accounting for false positives, we are able to make full use of the data from even our deepest nights, whose faintest detections cannot be recovered in another epoch. Additionally, our statistics remain reliable near the detection limit where false positives tend to accumulate.

Using $554.4$ single-night detections from $6$ unique DECam pointings, we computed the luminosity function of the Kuiper Belt as a whole (Section \ref{sec:luminosity}). Then using $\sim 280.0$ CC detections from $5$ unique DECam pointings we calculated the luminosity function of the CC population (Appendix \ref{sec:cc-luminosity}), down to $m_r \gtrsim 26.5$. In both cases, we were able to confirm that a single power law is not suitable to describe the underlying distribution. We found that rolling, broken, and exponentially tapered power laws yielded acceptable fits, though low discrimination power at the bright end prevented us from giving overwhelming preference to any model.

The most significant scientific result of this work is a measurement of the absolute magnitude distribution of the CCs down to $H_r \sim 10.5$. While a dearth of bright objects limits our constraining power at the bright end, our plethora of faint detections enable us to tightly constrain the faint end of the distribution. Our detections are consistent with an exponentially tapered power law with a faint-end slope $\alpha = 0.26^{+0.08}_{-0.07}$. This faint-end slope is marginally shallower than previous measurements in the literature, but is consistent with simulations of planetesimal formation via the streaming instability. This is of particular interest because the theory predicts a physically motivated size distribution (as opposed to other size distributions which have simply been engineered to adequately describe observations) of CCs that matches observations over the full range of sizes probed to date. However, we urge that these conclusions must be approached with caution. Of particular consequence to the conclusions from this work, limited resolution in SI simulations leaves the theoretical faint-end slope somewhat uncertain (see \citet{Kavelaars2021} for an in-depth discussion of the outstanding problems with the streaming instability as a complete theory of planetesimal formation). While the exponentially tapered power law $H$ distribution is a good fit to our CC detections, we cannot rule out rolling or broken power laws. Both the exponentially tapered and rolling power law distributions are consistent with the results of \citet{bernstein2004}, implying that the $H$ distribution continues to flatten out beyond the DEEP limit. In contrast, our broken power law may be in tension with the the results of \citet{bernstein2004}. Interestingly, our rolling power law fit would imply a characteristic size for the CCs, beyond which the probability density rolls over.

Finally, we note some limitations of this work that can be improved upon in future studies. The most obvious limitation is our use of single-night detections. Although we developed robust new techniques to account for false positives, linked detections with well-constrained orbits are still preferable. Since we used single-night detections in this work, our selection of a CC subsample of our detections was rather rough, relying heavily on the OSSOS\texttt{++} solar system model and the serendipity that Neptune was near these fields, meaning that they were necessarily relatively devoid of resonant objects. Linked orbits will enable proper dynamical classification, thus reducing the uncertainties in our studies of individual dynamical classes. In forthcoming work, we will analyze three additional fields similar to the B1 field. We will link our detections, yielding well-determined orbits for the smallest known KBOs. Our catalog of discoveries will enable studies of Kuiper Belt populations to unprecedented depth, providing deep insight to the formation and evolution of our planetary system.

\newpage
\appendix

\section{The Luminosity Function of the Cold Classicals}
\label{sec:cc-luminosity}
\begin{figure}[h]
    \centering
    \includegraphics[width=1\textwidth]{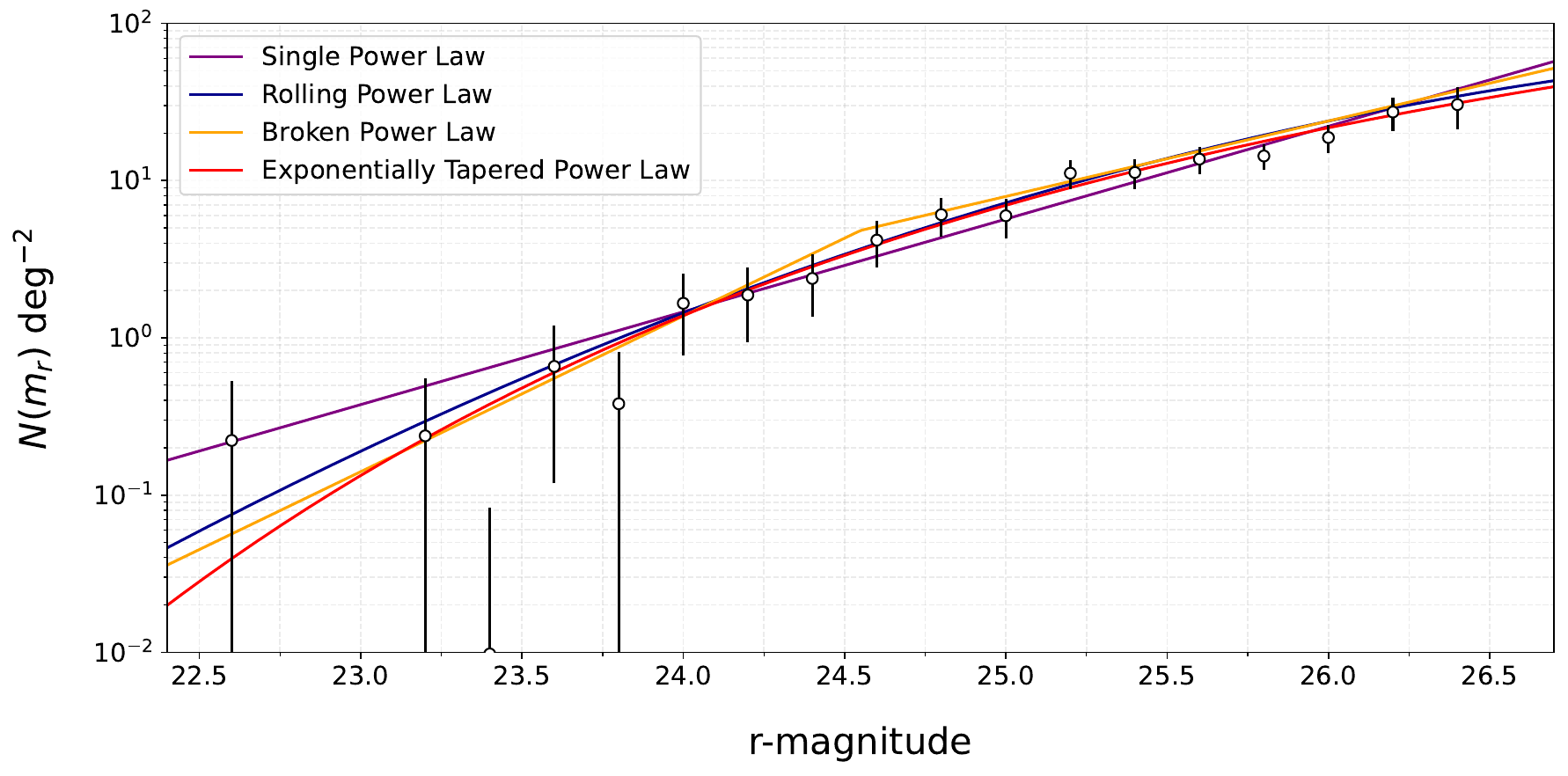}
    \caption{Best-fit differential distributions for the single, rolling, broken, and exponentially tapered power laws in purple, blue, yellow, and red, respectively. The points and 1-$\sigma$ error bars represent the differential distribution of our detections, corrected for efficiency and weight. Note that our fits are not normalized with respect to latitude, but rather they represent the average sky density among the fields in our pointing mosaic.}
    \label{fig:cc-luminosity}
\end{figure}

Here we calculate the luminosity function the subset of our detections that have rates consistent with being CCs (see Section \ref{sec:dynamics}). For these fits, we follow the same general sampling procedure as Section \ref{sec:H}. We summarize our results in Table \ref{tab:cc-luminosity} and show the fits in Figure \ref{fig:cc-luminosity}.

\begin{table}[h]
\renewcommand{\arraystretch}{1.25}
\centering
\begin{tabular*}{1\textwidth}{l@{\extracolsep{\fill}}lccc}
\hline
\hline
Functional Form & Best-Fit Parameters & $\langle \text{BIC} \rangle$ & $\langle Q_{outlier} \rangle$\\ \hline
    Single Power Law & $\langle \alpha, m_0\rangle = \langle 0.59^{+0.04}_{-0.04}, 23.72^{+0.12}_{-0.13} \rangle$ & $1.9$ & $0.93$ \\ 
    \hline
    Rolling Power Law & $\langle \alpha_1, \alpha_2, \Sigma_{23}\rangle = \langle 0.97^{+0.20}_{-0.17}, -0.09^{+0.04}_{-0.04}, 0.19^{+0.10}_{-0.07} \rangle$ & $0$ & $0.21$ \\ 
    \hline
    Broken Power Law & $\langle \alpha_1, \alpha_2, m_0, m_B\rangle = \langle 0.99^{+0.43}_{-0.21}, 0.48^{+0.07}_{-0.08}, 23.86^{+0.14}_{-0.18}, 24.55^{+0.43}_{-0.66} \rangle$ & $3.6$ & $0.55$ \\
    \hline
    Tapered Power Law & $\langle \alpha, \beta, m_0, m_B\rangle = \langle 0.16^{+0.09}_{-0.05}, 0.17^{+0.07}_{-0.05}, 12.23^{+5.73}_{-8.05}, 28.02^{+2.47}_{-1.50} \rangle$ & $7.9$ & $0.80$ \\
\hline
\end{tabular*}
\caption{Best fit parameters and statistics for each of the distributions tested for the luminosity function of the CC subsample of our detections. The BIC values have each been rescaled such that the minimum value among the distributions is 0.}
\label{tab:cc-luminosity}
\end{table}

Here a single power law provides a very marginal fit. As with the full set of KBOs, we cannot rule out the rolling, broken, or exponentially tapered power laws.

\acknowledgements
We thank Gary Bernstein and an anonymous referee for insightful and engaged reviews that have improved the clarity and fidelity of this paper.

This work is based in part on observations at Cerro Tololo Inter-American Observatory at NSF’s NOIRLab (NOIRLab Prop. ID 2019A-0337; PI: D. Trilling), which is managed by the Association of Universities for Research in Astronomy (AURA) under a cooperative agreement with the National Science Foundation.

This material is based upon work supported by the National Aeronautics and Space Administration under grant No.\ NNX17AF21G issued through the SSO Planetary Astronomy Program and by the National Science Foundation under grant No.\ AST-2009096. This research was supported in part through computational resources and services provided by Advanced Research Computing at the University of Michigan, Ann Arbor.

\bibliographystyle{aasjournal}
\bibliography{references}

\begin{thebibliography}{}
\expandafter\ifx\csname natexlab\endcsname\relax\def\natexlab#1{#1}\fi
\providecommand{\url}[1]{\href{#1}{#1}}
\providecommand{\dodoi}[1]{doi:~\href{http://doi.org/#1}{\nolinkurl{#1}}}
\providecommand{\doeprint}[1]{\href{http://ascl.net/#1}{\nolinkurl{http://ascl.net/#1}}}
\providecommand{\doarXiv}[1]{\href{https://arxiv.org/abs/#1}{\nolinkurl{https://arxiv.org/abs/#1}}}

\bibitem[{Abadi {et~al.}(2015)}]{tensorflow}
Abadi, M., {et~al.} 2015, {TensorFlow}: Large-Scale Machine Learning on Heterogeneous Systems.
\newblock \url{https://www.tensorflow.org/}

\bibitem[{{Abod} {et~al.}(2019){Abod}, {Simon}, {Li}, {Armitage}, {Youdin}, \& {Kretke}}]{Abod2019}
{Abod}, C.~P., {Simon}, J.~B., {Li}, R., {et~al.} 2019, \apj, 883, 192, \dodoi{10.3847/1538-4357/ab40a3}

\bibitem[{{Alard} \& {Lupton}(1998)}]{Alard1998}
{Alard}, C., \& {Lupton}, R.~H. 1998, \apj, 503, 325, \dodoi{10.1086/305984}

\bibitem[{{Allen} {et~al.}(2001){Allen}, {Bernstein}, \& {Malhotra}}]{DT2}
{Allen}, R.~L., {Bernstein}, G.~M., \& {Malhotra}, R. 2001, The Astrophysical Journal Letters, 549

\bibitem[{{Bannister} {et~al.}(2018){Bannister}, {Gladman}, {Kavelaars}, {Petit}, {Volk}, {Chen}, {Alexandersen}, {Gwyn}, {Schwamb}, {Ashton}, {Benecchi}, {Cabral}, {Dawson}, {Delsanti}, {Fraser}, {Granvik}, {Greenstreet}, {Guilbert-Lepoutre}, {Ip}, {Jakubik}, {Jones}, {Kaib}, {Lacerda}, {Van Laerhoven}, {Lawler}, {Lehner}, {Lin}, {Lykawka}, {Marsset}, {Murray-Clay}, {Pike}, {Rousselot}, {Shankman}, {Thirouin}, {Vernazza}, \& {Wang}}]{OSSOSVII}
{Bannister}, M.~T., {Gladman}, B.~J., {Kavelaars}, J.~J., {et~al.} 2018, \apjs, 236, 18, \dodoi{10.3847/1538-4365/aab77a}

\bibitem[{{Barbary}(2016)}]{sep}
{Barbary}, K. 2016, The Journal of Open Source Software, 1, \dodoi{10.21105/joss.00058}

\bibitem[{{Becker}(2015)}]{hotpants}
{Becker}, A. 2015, {HOTPANTS: High Order Transform of PSF ANd Template Subtraction}.
\newblock \doeprint{1504.004}

\bibitem[{{Bernardinelli} {et~al.}(2022){Bernardinelli}, {Bernstein}, {Sako}, {Yanny}, {Aguena}, {Allam}, {Andrade-Oliveira}, {Bertin}, {Brooks}, {Buckley-Geer}, {Burke}, {Rosell}, {Carrasco Kind}, {Carretero}, {Conselice}, {Costanzi}, {da Costa}, {De Vicente}, {Desai}, {Diehl}, {Dietrich}, {Doel}, {Eckert}, {Everett}, {Ferrero}, {Flaugher}, {Fosalba}, {Frieman}, {Garc{\'\i}a-Bellido}, {Gerdes}, {Gruen}, {Gruendl}, {Gschwend}, {Hinton}, {Hollowood}, {Honscheid}, {James}, {Kent}, {Kuehn}, {Kuropatkin}, {Lahav}, {Maia}, {March}, {Menanteau}, {Miquel}, {Morgan}, {Myles}, {Ogando}, {Palmese}, {Paz-Chinch{\'o}n}, {Pieres}, {Malag{\'o}n}, {Romer}, {Roodman}, {Sanchez}, {Scarpine}, {Schubnell}, {Serrano}, {Sevilla-Noarbe}, {Smith}, {Soares-Santos}, {Suchyta}, {Swanson}, {Tarle}, {To}, {Varga}, \& {Walker}}]{bernardinelli2022}
{Bernardinelli}, P.~H., {Bernstein}, G.~M., {Sako}, M., {et~al.} 2022, \apjs, 258, 41, \dodoi{10.3847/1538-4365/ac3914}

\bibitem[{{Bernstein} {et~al.}(2004){Bernstein}, {Trilling}, {Allen}, {Brown}, {Holman}, \& {Malhotra}}]{bernstein2004}
{Bernstein}, G.~M., {Trilling}, D.~E., {Allen}, R.~L., {et~al.} 2004, \aj, 128, 1364, \dodoi{10.1086/422919}

\bibitem[{{Bertin} \& {Arnouts}(1996)}]{SourceExtractor}
{Bertin}, E., \& {Arnouts}, S. 1996, Astronomy and Astrophysics, Supplement, 117, \dodoi{10.1051/aas:1996164}

\bibitem[{{Bertin} {et~al.}(2002)}]{SWarp}
{Bertin}, E., {et~al.} 2002, in Astronomical Society of the Pacific Conference Series, Vol. 281, Astronomical Data Analysis Software and Systems XI, ed. D.~A. {Bohlender}, D.~{Durand}, \& T.~H. {Handley}, 228

\bibitem[{{Crompvoets} {et~al.}(2022){Crompvoets}, {Lawler}, {Volk}, {Chen}, {Gladman}, {Peltier}, {Alexandersen}, {Bannister}, {Gwyn}, {Kavelaars}, \& {Petit}}]{Crompvoets2022}
{Crompvoets}, B.~L., {Lawler}, S.~M., {Volk}, K., {et~al.} 2022, PSJ, 3, 113, \dodoi{10.3847/PSJ/ac67e0}

\bibitem[{Ester {et~al.}(1996)Ester, Kriegel, Sander, \& Xu}]{dbscan}
Ester, M., Kriegel, H.-P., Sander, J., \& Xu, X. 1996, in Proceedings of the Second International Conference on Knowledge Discovery and Data Mining, KDD'96 (AAAI Press), 226–231

\bibitem[{{Fraser} {et~al.}(2014){Fraser}, {Brown}, {Morbidelli}, {Parker}, \& {Batygin}}]{fraser2014}
{Fraser}, W.~C., {Brown}, M.~E., {Morbidelli}, A., {Parker}, A., \& {Batygin}, K. 2014, \apj, 782, 100, \dodoi{10.1088/0004-637X/782/2/100}

\bibitem[{{Fraser} \& {Kavelaars}(2009)}]{Fraser2009}
{Fraser}, W.~C., \& {Kavelaars}, J.~J. 2009, \aj, 137, 72, \dodoi{10.1088/0004-6256/137/1/72}

\bibitem[{{Fraser} {et~al.}(2008){Fraser}, {Kavelaars}, {Holman}, {Pritchet}, {Gladman}, {Grav}, {Jones}, {MacWilliams}, \& {Petit}}]{Fraser2008}
{Fraser}, W.~C., {Kavelaars}, J.~J., {Holman}, M.~J., {et~al.} 2008, \icarus, 195, 827, \dodoi{10.1016/j.icarus.2008.01.014}

\bibitem[{{Fuentes} {et~al.}(2009){Fuentes}, {George}, \& {Holman}}]{Fuentes2009}
{Fuentes}, C.~I., {George}, M.~R., \& {Holman}, M.~J. 2009, \apj, 696, 91, \dodoi{10.1088/0004-637X/696/1/91}

\bibitem[{{Gladman} \& {Kavelaars}(1997)}]{DT1}
{Gladman}, B., \& {Kavelaars}, J.~J. 1997, Astronomy and Astrophysics, 317

\bibitem[{{Gladman} \& {Volk}(2021)}]{GladmanVolk2021}
{Gladman}, B., \& {Volk}, K. 2021, \araa, 59, 203, \dodoi{10.1146/annurev-astro-120920-010005}

\bibitem[{{Goldstein} {et~al.}(2015)}]{autoscan}
{Goldstein}, D.~A., {et~al.} 2015, The Astronomical Journal, 150, \dodoi{10.1088/0004-6256/150/3/82}

\bibitem[{{Heinze} {et~al.}(2015){Heinze}, {Metchev}, \& {Trollo}}]{DT5}
{Heinze}, A.~N., {Metchev}, S., \& {Trollo}, J. 2015, The Astronomical Journal, 150

\bibitem[{{Holman} {et~al.}(2004)}]{DT4}
{Holman}, M.~J., {et~al.} 2004, Nature, 430

\bibitem[{{Ivezi{\'c}} {et~al.}(2019){Ivezi{\'c}}, {Kahn}, {Tyson}, {Abel}, {Acosta}, {Allsman}, {Alonso}, {AlSayyad}, {Anderson}, {Andrew}, {Angel}, {Angeli}, {Ansari}, {Antilogus}, {Araujo}, {Armstrong}, {Arndt}, {Astier}, {Aubourg}, {Auza}, {Axelrod}, {Bard}, {Barr}, {Barrau}, {Bartlett}, {Bauer}, {Bauman}, {Baumont}, {Bechtol}, {Bechtol}, {Becker}, {Becla}, {Beldica}, {Bellavia}, {Bianco}, {Biswas}, {Blanc}, {Blazek}, {Blandford}, {Bloom}, {Bogart}, {Bond}, {Booth}, {Borgland}, {Borne}, {Bosch}, {Boutigny}, {Brackett}, {Bradshaw}, {Brandt}, {Brown}, {Bullock}, {Burchat}, {Burke}, {Cagnoli}, {Calabrese}, {Callahan}, {Callen}, {Carlin}, {Carlson}, {Chandrasekharan}, {Charles-Emerson}, {Chesley}, {Cheu}, {Chiang}, {Chiang}, {Chirino}, {Chow}, {Ciardi}, {Claver}, {Cohen-Tanugi}, {Cockrum}, {Coles}, {Connolly}, {Cook}, {Cooray}, {Covey}, {Cribbs}, {Cui}, {Cutri}, {Daly}, {Daniel}, {Daruich}, {Daubard}, {Daues}, {Dawson}, {Delgado}, {Dellapenna}, {de Peyster}, {de Val-Borro}, {Digel}, {Doherty}, {Dubois},
  {Dubois-Felsmann}, {Durech}, {Economou}, {Eifler}, {Eracleous}, {Emmons}, {Fausti Neto}, {Ferguson}, {Figueroa}, {Fisher-Levine}, {Focke}, {Foss}, {Frank}, {Freemon}, {Gangler}, {Gawiser}, {Geary}, {Gee}, {Geha}, {Gessner}, {Gibson}, {Gilmore}, {Glanzman}, {Glick}, {Goldina}, {Goldstein}, {Goodenow}, {Graham}, {Gressler}, {Gris}, {Guy}, {Guyonnet}, {Haller}, {Harris}, {Hascall}, {Haupt}, {Hernandez}, {Herrmann}, {Hileman}, {Hoblitt}, {Hodgson}, {Hogan}, {Howard}, {Huang}, {Huffer}, {Ingraham}, {Innes}, {Jacoby}, {Jain}, {Jammes}, {Jee}, {Jenness}, {Jernigan}, {Jevremovi{\'c}}, {Johns}, {Johnson}, {Johnson}, {Jones}, {Juramy-Gilles}, {Juri{\'c}}, {Kalirai}, {Kallivayalil}, {Kalmbach}, {Kantor}, {Karst}, {Kasliwal}, {Kelly}, {Kessler}, {Kinnison}, {Kirkby}, {Knox}, {Kotov}, {Krabbendam}, {Krughoff}, {Kub{\'a}nek}, {Kuczewski}, {Kulkarni}, {Ku}, {Kurita}, {Lage}, {Lambert}, {Lange}, {Langton}, {Le Guillou}, {Levine}, {Liang}, {Lim}, {Lintott}, {Long}, {Lopez}, {Lotz}, {Lupton}, {Lust}, {MacArthur}, {Mahabal},
  {Mandelbaum}, {Markiewicz}, {Marsh}, {Marshall}, {Marshall}, {May}, {McKercher}, {McQueen}, {Meyers}, {Migliore}, {Miller}, {Mills}, {Miraval}, {Moeyens}, {Moolekamp}, {Monet}, {Moniez}, {Monkewitz}, {Montgomery}, {Morrison}, {Mueller}, {Muller}, {Mu{\~n}oz Arancibia}, {Neill}, {Newbry}, {Nief}, {Nomerotski}, {Nordby}, {O'Connor}, {Oliver}, {Olivier}, {Olsen}, {O'Mullane}, {Ortiz}, {Osier}, {Owen}, {Pain}, {Palecek}, {Parejko}, {Parsons}, {Pease}, {Peterson}, {Peterson}, {Petravick}, {Libby Petrick}, {Petry}, {Pierfederici}, {Pietrowicz}, {Pike}, {Pinto}, {Plante}, {Plate}, {Plutchak}, {Price}, {Prouza}, {Radeka}, {Rajagopal}, {Rasmussen}, {Regnault}, {Reil}, {Reiss}, {Reuter}, {Ridgway}, {Riot}, {Ritz}, {Robinson}, {Roby}, {Roodman}, {Rosing}, {Roucelle}, {Rumore}, {Russo}, {Saha}, {Sassolas}, {Schalk}, {Schellart}, {Schindler}, {Schmidt}, {Schneider}, {Schneider}, {Schoening}, {Schumacher}, {Schwamb}, {Sebag}, {Selvy}, {Sembroski}, {Seppala}, {Serio}, {Serrano}, {Shaw}, {Shipsey}, {Sick}, {Silvestri},
  {Slater}, {Smith}, {Smith}, {Sobhani}, {Soldahl}, {Storrie-Lombardi}, {Stover}, {Strauss}, {Street}, {Stubbs}, {Sullivan}, {Sweeney}, {Swinbank}, {Szalay}, {Takacs}, {Tether}, {Thaler}, {Thayer}, {Thomas}, {Thornton}, {Thukral}, {Tice}, {Trilling}, {Turri}, {Van Berg}, {Vanden Berk}, {Vetter}, {Virieux}, {Vucina}, {Wahl}, {Walkowicz}, {Walsh}, {Walter}, {Wang}, {Wang}, {Warner}, {Wiecha}, {Willman}, {Winters}, {Wittman}, {Wolff}, {Wood-Vasey}, {Wu}, {Xin}, {Yoachim}, \& {Zhan}}]{LSST}
{Ivezi{\'c}}, {\v{Z}}., {Kahn}, S.~M., {Tyson}, J.~A., {et~al.} 2019, \apj, 873, 111, \dodoi{10.3847/1538-4357/ab042c}

\bibitem[{{Jewitt} \& {Luu}(1993)}]{jewitt1993}
{Jewitt}, D., \& {Luu}, J. 1993, \nat, 362, 730, \dodoi{10.1038/362730a0}

\bibitem[{{Kavelaars} {et~al.}(2021)}]{Kavelaars2021}
{Kavelaars}, J.~J., {et~al.} 2021, Astrophysical Journal Letters, 920, \dodoi{10.3847/2041-8213/ac2c72}

\bibitem[{{Kessler} {et~al.}(2015){Kessler}, {Marriner}, {Childress}, {Covarrubias}, {D'Andrea}, {Finley}, {Fischer}, {Foley}, {Goldstein}, {Gupta}, {Kuehn}, {Marcha}, {Nichol}, {Papadopoulos}, {Sako}, {Scolnic}, {Smith}, {Sullivan}, {Wester}, {Yuan}, {Abbott}, {Abdalla}, {Allam}, {Benoit-L{\'e}vy}, {Bernstein}, {Bertin}, {Brooks}, {Carnero Rosell}, {Carrasco Kind}, {Castander}, {Crocce}, {da Costa}, {Desai}, {Diehl}, {Eifler}, {Fausti Neto}, {Flaugher}, {Frieman}, {Gerdes}, {Gruen}, {Gruendl}, {Honscheid}, {James}, {Kuropatkin}, {Li}, {Maia}, {Marshall}, {Martini}, {Miller}, {Miquel}, {Nord}, {Ogando}, {Plazas}, {Reil}, {Romer}, {Roodman}, {Sanchez}, {Sevilla-Noarbe}, {Smith}, {Soares-Santos}, {Sobreira}, {Tarle}, {Thaler}, {Thomas}, {Tucker}, {Walker}, \& {DES Collaboration}}]{Kessler2015}
{Kessler}, R., {Marriner}, J., {Childress}, M., {et~al.} 2015, \aj, 150, 172, \dodoi{10.1088/0004-6256/150/6/172}

\bibitem[{{Lawler} {et~al.}(2018)}]{OSSOSVIII}
{Lawler}, S.~M., {et~al.} 2018, The Astronomical Journal, 155, \dodoi{10.3847/1538-3881/aab8ff}

\bibitem[{{Loredo}(2004)}]{Loredo2004}
{Loredo}, T.~J. 2004, in American Institute of Physics Conference Series, Vol. 735, Bayesian Inference and Maximum Entropy Methods in Science and Engineering: 24th International Workshop on Bayesian Inference and Maximum Entropy Methods in Science and Engineering, ed. R.~{Fischer}, R.~{Preuss}, \& U.~V. {Toussaint}, 195--206, \dodoi{10.1063/1.1835214}

\bibitem[{{Magnier} {et~al.}(2013){Magnier}, {Schlafly}, {Finkbeiner}, {Juric}, {Tonry}, {Burgett}, {Chambers}, {Flewelling}, {Kaiser}, {Kudritzki}, {Morgan}, {Price}, {Sweeney}, \& {Stubbs}}]{PS1_Photometry2}
{Magnier}, E.~A., {Schlafly}, E., {Finkbeiner}, D., {et~al.} 2013, \apjs, 205, 20, \dodoi{10.1088/0067-0049/205/2/20}

\bibitem[{Manski \& Lerman(1977)}]{weighted-likelihood}
Manski, C., \& Lerman, S.~R. 1977, Econometrica, 45, 1977.
\newblock \url{https://EconPapers.repec.org/RePEc:ecm:emetrp:v:45:y:1977:i:8:p:1977-88}

\bibitem[{{Millis} {et~al.}(2002){Millis}, {Buie}, {Wasserman}, {Elliot}, {Kern}, \& {Wagner}}]{deep-ecliptic-survey}
{Millis}, R.~L., {Buie}, M.~W., {Wasserman}, L.~H., {et~al.} 2002, \aj, 123, 2083, \dodoi{10.1086/339481}

\bibitem[{Napier(2020)}]{Napier_spacerock}
Napier, K. 2020, SpaceRocks, \url{https://github.com/kjnapier/spacerocks}

\bibitem[{{Nesvorn{\'y}}(2018)}]{Nesvorny2018}
{Nesvorn{\'y}}, D. 2018, \araa, 56, 137, \dodoi{10.1146/annurev-astro-081817-052028}

\bibitem[{{Parker} \& {Kavelaars}(2010)}]{Parker2010}
{Parker}, A.~H., \& {Kavelaars}, J.~J. 2010, Publications of the Astronomical Society of the Pacific, 122, \dodoi{10.1086/652424}

\bibitem[{{Petit} {et~al.}(2006){Petit}, {Holman}, {Gladman}, {Kavelaars}, {Scholl}, \& {Loredo}}]{Petit2006}
{Petit}, J.~M., {Holman}, M.~J., {Gladman}, B.~J., {et~al.} 2006, \mnras, 365, 429, \dodoi{10.1111/j.1365-2966.2005.09661.x}

\bibitem[{{Petit} {et~al.}(2011){Petit}, {Kavelaars}, {Gladman}, {Jones}, {Parker}, {Van Laerhoven}, {Nicholson}, {Mars}, {Rousselot}, {Mousis}, {Marsden}, {Bieryla}, {Taylor}, {Ashby}, {Benavidez}, {Campo Bagatin}, \& {Bernabeu}}]{CFEPSPetit2011}
{Petit}, J.~M., {Kavelaars}, J.~J., {Gladman}, B.~J., {et~al.} 2011, \aj, 142, 131, \dodoi{10.1088/0004-6256/142/4/131}

\bibitem[{{Stansberry} {et~al.}(2021){Stansberry}, {Fraser}, {Trilling}, {Bernstein}, {Grundy}, \& {Holman}}]{stansberry2021}
{Stansberry}, J.~A., {Fraser}, W.~C., {Trilling}, D.~E., {et~al.} 2021, {An Ultra-Sensitive Pencil Beam Search for 10 km Trans-Neptunian Objects}, JWST Proposal. Cycle 1, ID. \#1568

\bibitem[{{Tonry} {et~al.}(2012){Tonry}, {Stubbs}, {Lykke}, {Doherty}, {Shivvers}, {Burgett}, {Chambers}, {Hodapp}, {Kaiser}, {Kudritzki}, {Magnier}, {Morgan}, {Price}, \& {Wainscoat}}]{PS1_Photometry1}
{Tonry}, J.~L., {Stubbs}, C.~W., {Lykke}, K.~R., {et~al.} 2012, \apj, 750, 99, \dodoi{10.1088/0004-637X/750/2/99}

\bibitem[{{Trilling} {et~al.}(2023)}]{deepi}
{Trilling}, D.~E., {et~al.} 2023, Submitted to AJ

\bibitem[{{Trujillo} {et~al.}(2023)}]{deepii}
{Trujillo}, C.~A., {et~al.} 2023, Submitted to AJ

\bibitem[{{Tyson} {et~al.}(1992){Tyson}, {Guhathakurta}, {Bernstein}, \& {Hut}}]{1992AAS...181.0610T}
{Tyson}, J.~A., {Guhathakurta}, P., {Bernstein}, G.~M., \& {Hut}, P. 1992, in American Astronomical Society Meeting Abstracts, Vol. 181, American Astronomical Society Meeting Abstracts, 06.10

\bibitem[{{Valdes} {et~al.}(2014){Valdes}, {Gruendl}, \& {DES Project}}]{CommunityPipeline}
{Valdes}, F., {Gruendl}, R., \& {DES Project}. 2014, in Astronomical Society of the Pacific Conference Series, Vol. 485, Astronomical Data Analysis Software and Systems XXIII, ed. N.~{Manset} \& P.~{Forshay}, 379

\bibitem[{{Whidden} {et~al.}(2019){Whidden}, {Bryce Kalmbach}, {Connolly}, {Jones}, {Smotherman}, {Bektesevic}, {Slater}, {Becker}, {Ivezi{\'c}}, {Juri{\'c}}, {Bolin}, {Moeyens}, {F{\"o}rster}, \& {Golkhou}}]{kbmod}
{Whidden}, P.~J., {Bryce Kalmbach}, J., {Connolly}, A.~J., {et~al.} 2019, \aj, 157, 119, \dodoi{10.3847/1538-3881/aafd2d}

\end{thebibliography}

\end{document}